\documentclass[aps,prl,nofootinbib,twocolumn,showpacs,superscriptaddress,letterpaper,amsmath,amssymb]{revtex4}
\usepackage{graphicx}  
\usepackage{dcolumn}   
\usepackage{bm}        
\usepackage{amssymb}   
\usepackage{feynmf}
\usepackage{slashed}
\usepackage{multirow}
\usepackage{color}
\usepackage{array}

\newcolumntype{L}[1]{>{\raggedright\let\newline\\\arraybackslash\hspace{0pt}}m{#1}}
\newcolumntype{C}[1]{>{\centering\let\newline\\\arraybackslash\hspace{0pt}}m{#1}}
\newcolumntype{R}[1]{>{\raggedleft\let\newline\\\arraybackslash\hspace{0pt}}m{#1}}

\def\gsim{\lower0.5ex\hbox{$\:\buildrel >\over\sim\:$}}
\def\lsim{\lower0.5ex\hbox{$\:\buildrel <\over\sim\:$}}

\newcommand{\be}{\begin{equation}}
\newcommand{\ee}{\end{equation}}
\newcommand{\bea}{\begin{eqnarray}}
\newcommand{\eea}{\end{eqnarray}}

\newcommand{\nbox}{{\,\lower0.9pt\vbox{\hrule \hbox{\vrule height 0.2 cm
\hskip 0.2 cm \vrule height 0.2 cm}\hrule}\,}}

\newcommand{\GeV}{\text{GeV}}

\newcommand{\Dslash}{\slashed{D}}

\def\sub#1{_{\lower.25ex\hbox{$\scriptstyle#1$}}}

\def\to{\rightarrow}

\newskip\zatskip \zatskip=0pt plus0pt minus0pt
\def\matth{\mathsurround=0pt}
\def\lsim{\mathrel{\mathpalette\atversim<}}
\def\gsim{\mathrel{\mathpalette\atversim>}}
\def\sigv{\ifmmode \langle\sigma v\rangle\else $\langle\sigma v\rangle$\fi}
\newskip\zatskip \zatskip=0pt plus0pt minus0pt
\def\matth{\mathsurround=0pt}
\def\lsim{\mathrel{\mathpalette\atversim<}}
\def\gsim{\mathrel{\mathpalette\atversim>}}
\def\atversim#1#2{\lower0.7ex\vbox{\baselineskip\zatskip\lineskip\zatskip
  \lineskiplimit
  0pt\ialign{$\matth#1\hfil##\hfil$\crcr#2\crcr\sim\crcr}}}

\def\missET {{\not\!\! E_{\textrm{T}}}}

\begin{document}

\thispagestyle{empty}
\vspace*{-3.5cm}

\vspace{0.5in}

\title{mono-$Z'$: searches for dark matter in events with a resonance and missing transverse energy}

\begin{center}
\begin{abstract}
We analyze the potential dark matter implications of LHC events with missing transverse momentum and a resonance, such as a  $Z'$, decaying to a pair of jets or leptons. This final state contains significant discovery potential, yet has not yet been examined in detail by the LHC experiments.  We introduce models of $Z'$ production in association with dark matter particles, propose reconstruction and selection strategies, and estimate the sensitivity of the current LHC dataset.
\end{abstract}
\end{center}

\author{Marcelo Autran}
\affiliation{Department of Physics and Astronomy, University of
  California, Irvine, CA 92697}
\author{Kevin Bauer}
\affiliation{Department of Physics and Astronomy, University of
  California, Irvine, CA 92697}
\author{Tongyan Lin}
\affiliation{Kavli Institute for Cosmological Physics and the Enrico
Fermi Institute, The University of Chicago, 5640 S. Ellis Ave,
Chicago, Il 60637}
\author{Daniel Whiteson}
\affiliation{Department of Physics and Astronomy, University of
  California, Irvine, CA 92697}


\date{\today}

\maketitle


\section{Introduction}

As the Large Hadron Collider (LHC) resumes operations this year after a major upgrade and a half-decade of data taking, a central area of focus will be the search for physics beyond the standard model (SM).  While the LHC will have sensitivity to many models inspired by theoretical extensions or generalizations of the SM, the search for dark matter is of particular interest due to the well-established fact of its existence~\cite{Bertone:2004pz}. The collider detection of dark matter is a cornerstone of the effort to elucidate and obtain evidence for the particle nature of dark matter, and is complementary to astrophysical methods of detection.

Searches for dark matter production at the LHC rely on the production of a visible object $X$ recoiling against the missing transverse momentum ($\missET$) from the invisible dark matter particles. Cases where $X$ is a SM particle such as $g/q$~\cite{Beltran:2010ww,Fox:2011pm,Goodman:2010ku,Rajaraman:2011wf,Aad:2015zva,Khachatryan:2014rra}, $W$~\cite{Bai:2012xg,Aad:2013oja,ATLAS:2014wra,Khachatryan:2014tva}, $Z$~\cite{Carpenter:2012rg,Aad:2014vka},     $H$~\cite{Carpenter:2013xra,Berlin:2014cfa}, $\gamma$~\cite{Fox:2011pm,Khachatryan:2014rwa,Aad:2014tda}, or a heavy quark~\cite{Lin:2013sca,Haisch:2015ioa,CMS:2014pvf,Aad:2014vea} have been considered.  For a review of simplified models for dark matter at the LHC, see Refs.~\cite{Abdallah:2014hon,Malik:2014ggr}.

In this paper, we present a new mechanism for dark matter production at the LHC, where the visible object is itself a new particle, a $Z'$ boson. We propose examples of models giving rise to a signal of $Z'+\missET$, where the $Z'$ boson can decay to pairs of charged leptons ($\ell^+\ell^-$) or to pairs of quarks leading to jets ($jj$), and is therefore distinguishable as a resonance in the dilepton or dijet mass spectrum. In each case, we study the sensitivity of the LHC in this channel, and compare with existing searches for the $Z'$ without a requirement of large $\missET$.

The models here specifically target the production of a new $Z'$ which is present in a hypothetical, non-minimal dark sector. New $Z'$ bosons arise in many extensions to the SM~\cite{Langacker:2008yv}, and the possibility of dark matter coupled to a $Z'$ has been explored extensively in the literature, including in the context of the LHC (see, e.g.~\cite{Petriello:2008pu,Gershtein:2008bf,Fox:2011pm,An:2012va,An:2012ue,Frandsen:2012rk,Arcadi:2013qia,Alves:2013tqa,Busoni:2014sya,Alves:2015pea}). It should be noted that the experimental signature of a dijet or dilepton resonance plus missing transverse momentum does not require a $Z'$: other possibilities, including new scalar resonances or colored resonances, are natural directions to explore.

In addition to extending the current program of $X+\missET$ studies, the models presented here point to final states  whose LHC data remains unexamined and which are natural generalizations~\cite{Abdullah:2014oaa} of previously performed searches for $Z/W+\missET$ with $Z\rightarrow \ell\ell$ or $Z/W \rightarrow jj$.  These data therefore contain real, untapped discovery potential, independent of theoretical interest in models of dark matter involving $Z'$ bosons.

The models considered here are also examples of dark sector signals that, to some extent, could be hidden in existing $\missET$-based searches. Searching specifically for a dijet or dilepton resonance reduces the backgrounds and could give a strong hint of new physics. Furthermore, many searches have been optimized for new high-mass particles. For the examples below, we find that the most unconstrained parameter space is for relatively light $Z'$ states, those with $m_{Z'}$ below 100 GeV, where current LHC searches have low efficiency.

In the following, we first review experimental constraints on $Z'$ gauge bosons and then describe several models of $Z'+$ dark matter production. The range of $Z'$ mass explored is 50-800 GeV: for lower masses, dijet masses would be more difficult to reconstruct due to a smaller angular separation in the partons. Work on LHC signals of $Z' + \missET$  with lower values of $m_{Z'}$  will appear elsewhere~\cite{Lin:future}, while related work focusing on leptonic $Z'$ decays plus missing transverse momentum can be found in Ref.~\cite{Primulando:future}.

We consider two models with a minimal set of renormalizable interactions: dark-Higgsstrahlung from a $Z'$, with the dark Higgs decaying invisibly; and a dark sector with two states $\chi_{1,2}$ that couple off-diagonally to the $Z'$. We also study the case where the production of the dark-sector states is through a higher-dimension operator. We analyze the sensitivity of the current LHC run to these models in $jj+\missET$ and $\ell\ell+\missET$ final states, and compare to existing constraints.  For the renormalizable models, the $Z'  + \missET$ search has better sensitivity than direct resonance searches only for low $Z'$ masses. In the operator case, it is possible to probe the scale of new interactions to around a few TeV.

\section{Current Constraints on $Z'$ Bosons}
	
For simplicity, we assume a $U(1)'$  where the  $Z'$  has universal vector couplings to SM quarks:
\begin{equation}
	{\cal L} \supset - \sum_q g_q \bar q \gamma^\mu q Z'^\mu. \label{eq:Zprimequark}
\end{equation}
The couplings above are the same as for gauged baryon number $U(1)_B$ with $g_q = g_B/6$, where anomaly cancellation could be achieved with additional heavy quarks or with chiral matter in a dark sector. This possibility has been studied in detail in the context of dark matter ({\it e.g.},~\cite{Dulaney:2010dj,Graesser:2011vj}). However, we do not assume that the gauged baryon number is the origin of the $U(1)'$. For example, it is possible that the $Z'$ couplings to SM fermions are generated by higher-dimensional operators~\cite{Fox:2011qd} while the dark sector states are directly charged under the $Z'$.

When we consider dilepton searches, we will introduce additional free parameters for couplings of the $Z'$ to leptons. Since the production of the $Z'$ does not depend on the lepton coupling (except through the dependence on the $Z'$ width, which we neglect) we present constraints from dilepton resonances searches simply in terms of the $Z'$ branching ratio to the appropriate lepton flavor.

Although we will not impose any relationship between the $Z'$ coupling to quarks, leptons, or dark sector particles, one natural possibility is that of kinetic mixing~\cite{Holdom:1985ag}, where a mixing of $Z'$ and hypercharge generates couplings of the $Z'$ to SM fermions. Since the natural size of the couplings is small in this case ($10^{-2}$ or less), we do not consider this for the models that rely on dark matter production via the $Z'$ couplings to quarks. However, this gives a simple way for the $Z'$ to decay to visible states in our last model, where the $Z'$ is only produced in the decay of dark sector states.

The range of $Z'$ mass explored here is 50-800 GeV.  For heavier masses, constraints from dilepton or dijet resonance searches are precisely where LHC searches excel since backgrounds are relatively low. The $Z' + \missET$ signature has additional particles produced along with the $Z'$ and so has a smaller rate than direct $Z'$ production; therefore we expect it to be a less sensitive probe of the models in the high mass regime. Meanwhile, a low mass $Z'$ decaying to quarks is difficult to resolve as separate jets; however, it is possible that this regime could be studied by employing jet substructure techniques~\cite{Lin:future,Izaguirre:2014ira}.

\subsection{Dijet Constraints}

Direct dijet resonance searches constrain a $Z'$ coupling to quarks. We take limits on $g_q$ as a function of $M_{Z'}$ from Ref.~\cite{Dobrescu:2013cmh}, which compiles experimental results down to $M_{Z'} = 140$ GeV. Here, the lowest mass region was covered by UA2 \cite{Alitti:1993pn} with integrated luminosity of 10.9 pb$^{-1}$. At lower $Z'$ mass,  dijet resonances are more difficult to constrain due to the large QCD background.  Data on the dijet spectrum down to $m_{jj}= 48$ GeV have been published by UA2 \cite{Alitti:1990kw} (4.7 pb$^{-1}$) and down to 60 GeV from CDF \cite{Abe:1989gz} (26 nb$^{-1}$). While a reanalysis of the data would be needed to obtain limits on new resonances, we estimate that the UA2 dijet limits continue to weaken below 140 GeV, reaching $g_q \lesssim 1$ at $M_{Z'} =$ 50 GeV (see also \cite{Krnjaic:2011ub}).

Future LHC analyses may be able to provide more robust coverage of the low mass $M_{Z'}$ region.
This was studied in Ref.~\cite{An:2012ue}, which considered associated $Z'$ searches, such as a $Z'$ in addition to a $Z$, $\gamma$, or jet. Using the additional object in the final state to improve trigger efficiency, it was shown that LHC searches can have sensitivity even down to $M_{Z'} \approx 50$ GeV and couplings comparable to or better than the estimated UA2 dijet limits.
	
\subsection{Dilepton Constraints}

A $Z'$ coupling to electrons is strongly constrained by LEP measurements~\cite{Alcaraz:2006mx}. In the first two of our models, we will focus on the possibility that the $Z'$ has suppressed couplings to electrons but $O(1)$ branching ratio of the $Z'$ to muons (for example, see Ref.~\cite{Altmannshofer:2014cfa} and references therein).

Then if the $Z'$ has a preferred coupling to muons and to quarks, a hadron collider can give interesting limits relative to the LEP precision measurements. As a direct comparison to $Z' + \missET$ searches, we consider constraints from searches for dimuon resonances. Limits are available from the CDF collaboration~\cite{Aaltonen:2008ah}  down to $M_{Z'} = 100$ GeV, while ATLAS~\cite{Aad:2014cka} and CMS~\cite{Khachatryan:2014fba} limits extend down to $M_{Z'} = 150$ GeV and $M_{Z'} = 300$ GeV, respectively\footnote{LHC searches for the SM Higgs decay to dimuons can also be recast to place constraints down to $M_{Z'}$ = 110 GeV  \cite{ATLAS-CONF-2013-010,Khachatryan:2014aep}; we do not consider these analyses as they are not directly applicable to our models, and would not qualitatively change our conclusions.}. Here published results are not available below $M_{Z}$ due to the large Drell-Yan background.

Below the $Z$-pole, Ref.~\cite{Hoenig:2014dsa} showed that LHC measurements of the Drell-Yan spectrum at low invariant mass~\cite{Chatrchyan:2013tia} can be used to set strong constraints on a $Z'$ coupling to quarks and muons. The recast of the data leads to constraints on couplings at the $10^{-3}-10^{-2}$ level. (In the context of kinetic mixing, the current constraint is $\epsilon < 0.012$ and can reach $\epsilon = 5 \times 10^{-3}$ for a binned 8 TeV LHC analysis at $M_{Z'} = 50$ GeV.)

We also consider small, universal couplings of the $Z'$ to all of the charged leptons, as in the case of kinetic mixing. As discussed above, this will be most relevant in our third model (Inelastic EFT) where the $Z'$ may be very weakly coupled to SM fermions.

\subsection{Other Limits}

A light $Z'$ coupling to quarks contributes to the $Z$ hadronic width through $Z \to q \bar q Z' \to 4j$ and through a $Z\bar q q$ vertex correction~\cite{Carone:1994aa,Carone:1995pu}. Applying the results of Ref.~\cite{Carone:1994aa} to the most recent measurement of $R_Z = \Gamma(Z \to {\rm hadrons})/\Gamma(Z \to \mu^+ \mu^-)= 20.785 \pm 0.033$~\cite{Agashe:2014kda}, this places a constraint of $g_q \lesssim 0.4-0.6$ for $M_{Z'} =50-140$ GeV where there are no dijet resonance constraints.

Finally, even if a $Z'$ couples only to quarks, kinetic mixing of the $Z'$ with $Z, \gamma$ can be generated at one-loop. There are strong constraints on this kinetic mixing from precision electroweak measurements~\cite{Hook:2010tw,Carone:1995pu}, giving a bound $\epsilon \lesssim 0.02$ for $M_{Z'} \ll M_Z$. Since the kinetic mixing parameters are model-dependent, we do not examine this constraint any further, except to note that it is particularly strong for  $M_{Z'} \approx M_Z$ and so any model in this case would have to be particularly tuned.

\section{Models of $Z'+ \missET$ production}
	
\subsection{Dark Higgs}

\begin{figure}[t]
\begin{center}
 \includegraphics[width=\linewidth]{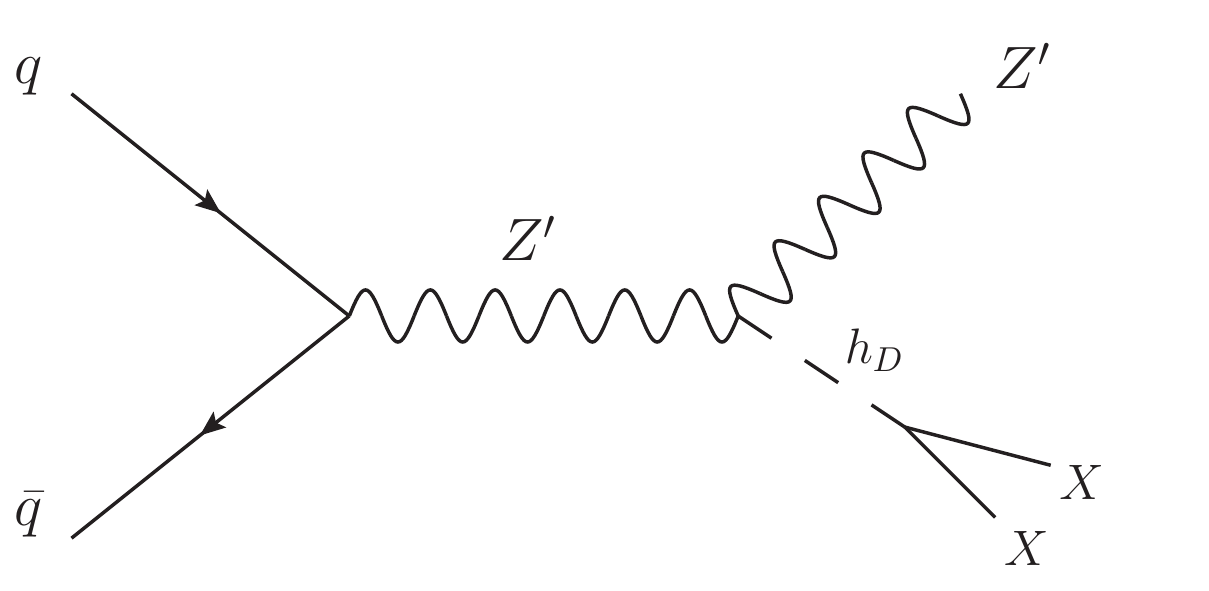}
 \caption{ Diagram of the production of a $Z'$ in association with a dark Higgs boson ($h_D$) which decays into two stable dark states, $\chi$.  It is assumed that $h_D$ is lighter than $2 M_{Z'}$ and decays with 100$\%$ branching to the invisible states. \label{fig:darkhiggs} }
 \end{center}
\end{figure}

A model with a new $Z'$ naturally comes with its own scalar (or set of scalars) responsible for spontaneous symmetry breaking. Suppose there is a new massive scalar that couples to the $Z'$, which we call the dark Higgs, $h_D$. Analogous to the SM process of Higgs-boson radiation from a $W$ or $Z$, the new scalar is radiated by the $Z'$ in a dark-Higgsstrahlung process. If this new dark Higgs boson additionally couples with invisible states\footnote{Another possibility is that the $Z'$ decays to dark matter, while the dark Higgs decays to SM states through mixing with the SM Higgs. Then the monojet search channel would also be sensitive to the model.}, its primary signature could be $\missET$, as shown in Fig.~\ref{fig:darkhiggs}.

As a minimal model for this process, we introduce a new $U(1)'$ with a charged scalar field $\Phi_D$ and an invisible singlet scalar $\phi_X$:
\begin{align}
	{\cal L} \supset & |D_\mu \Phi_D|^2 + \mu_D^2 |\Phi_D|^2 - \lambda_D |\Phi_D|^4  - \frac{1}{4} (F'_{\mu \nu})^2
	\nonumber \\
		& + \frac{1}{2} (\partial_\mu \phi_X)^2 - \lambda_X |\Phi_D|^2 \phi_X^2  - V(\phi_X).
\label{dhlag}
\end{align}
The dark Higgs field $\Phi_D = \frac{1}{\sqrt{2}} \left( v_D + h_D \right)$ obtains a vev $v_D$, giving mass to the $Z'$.  The masses of the dark scalars $h_D$ and $\phi_X$ are fixed by the scalar potentials, and the $Z'$ couplings to quarks are as in Eq.~\ref{eq:Zprimequark}.  Furthermore, if $m_X \gtrsim 100$ GeV or is very close to $m_h/2$, it is straightforward for $\phi_X$ to be a good thermal relic dark matter candidate if a scalar Higgs portal coupling is added to the Lagrangian in Eq.~\ref{dhlag}~\cite{deSimone:2014pda}. However, we do not require $\phi_X$ to be a thermal relic as this would impose a restriction on $M_{h_D}$.
	
The coupling of $h_D$ with the new gauge boson is
\begin{align}
	 Q_h g_z  M_{Z'} h_D Z'_\mu Z'^{\mu} \equiv g_{h_D} M_{Z'} h_D Z'_\mu Z'^{\mu},
\end{align}
where $Q_h$ is the charge of $\Phi_D$, which is a free parameter that we absorb by defining the effective coupling $g_{h_D}$. 
The dark Higgs can decay dominantly to the invisible $\phi_X$ states through the $\lambda_X$ coupling, which we can take to be $O(1)$.  Meanwhile, decays of $h_D \to Z' Z'^*$ will be suppressed as long as $m_{h_D} < 2 M_{Z'}$.  We assume the mixing of $h_D$ with the SM Higgs is small.
	
As discussed in the previous section, we will take the SM charges under the $Z'$ to be a separate free parameter, in order to be as general as possible. In considering signatures with dijets plus missing transverse momentum, we consider only the coupling to quarks; for dilepton plus missing transverse momentum signals, we will focus on the possibility of a non-zero branching fraction to muons.

The masses $M_{h_D}$ and $M_{Z'}$ are independent quantities in the model, though they are set by the same scale $v_D$, with $M_{h_D}/M_{Z'} = \sqrt{2}\lambda_D/g_{h_D}$. Note that since the $Z' + \missET$ signal due to the process shown in Fig.~\ref{fig:darkhiggs} favors larger $g_{h_D}$, this implies that for perturbative couplings the dark Higgs cannot be much heaver than the $Z'$. In order to capture most of the effects of different particle masses, we simply consider here two benchmark scenarios.
In the  ``light'' $M_{h_D}$ benchmark case, we set:
\begin{equation}
	M_{h_D} = \begin{cases}
		M_{Z'}	& ,\ M_{Z'} < 125\ \textrm{GeV} \\
		125\  \textrm{GeV}	& ,\ M_{Z'} > 125\ \textrm{GeV}, \\
	\end{cases}
\label{eq:darkhiggslight}
\end{equation}
In the  ``heavy'' $M_{h_D}$ benchmark case, we set\footnote{For the lowest mass point considered $M_{Z'}$ = 50 GeV, the decay of $h_D \rightarrow Z' Z'$ is kinematically allowed; for simplicity we continue to fix the $h_D$ invisible branching fraction to 1. 
}:
\begin{equation}
	M_{h_D} = \begin{cases}
		125\  \textrm{GeV}	& ,\ M_{Z'} < 125\ \textrm{GeV} \\
		M_{Z'}	& ,\ M_{h_D} > 125\ \textrm{GeV}. \\
	\end{cases}
\label{eq:darkhiggsheavy}
\end{equation}

\subsection{Light Vector}

\begin{figure}[t]
\includegraphics[width=0.97\linewidth]{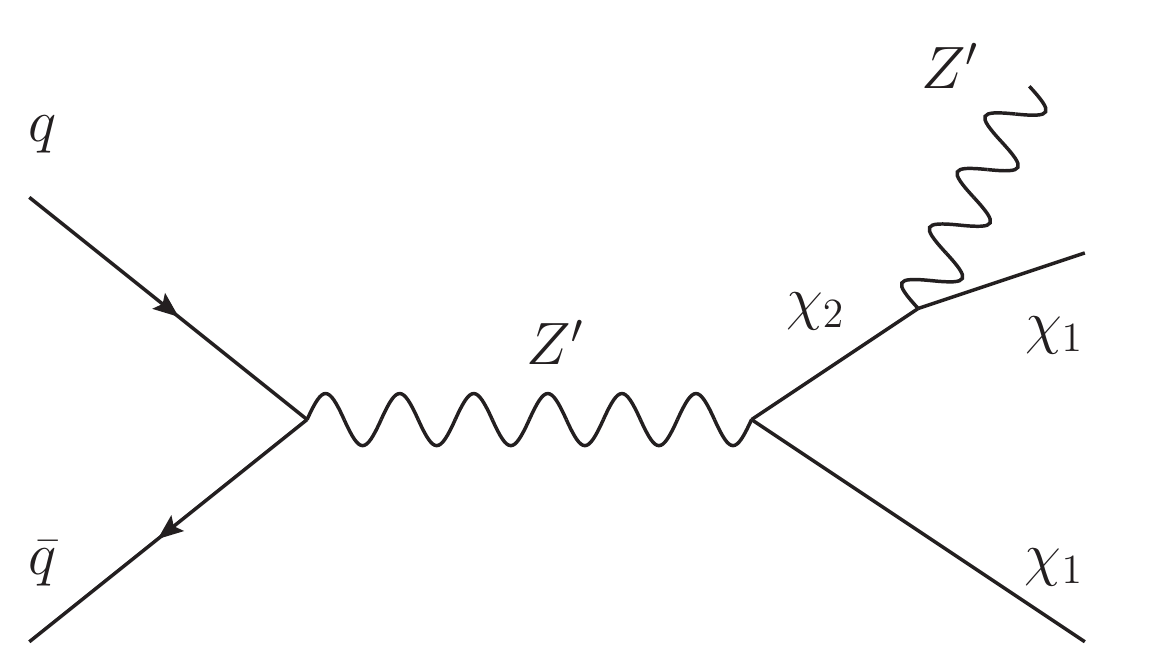}
\caption{ Diagram of the production of $\chi_1 \chi_2$, followed by decay of the heavier dark sector state $\chi_2$ to $Z' + \chi_1$, where $\chi_1$ is a possible dark matter candidate.}
\label{fig:light_feyn}
\end{figure}

When the $Z'$ is relatively light, it can be produced in the decays of dark sector states\footnote{Alternatively, the $Z'$ can be produced as radiation from off-shell dark sector states~\cite{Lin:future,Primulando:future}.}. An example is given in Fig.~\ref{fig:light_feyn}, where the $Z'$ possesses off-diagonal couplings to dark sector states $\chi_2$ and $\chi_1$. If the mass splitting between the two states is larger than $M_{Z'}$, the heavier state ($\chi_2$) can decay to an on-shell $Z'$ and a $\chi_1$. Meanwhile $\chi_1$ is stable and a dark matter candidate.

As a concrete example, we consider a $Z'$ coupled to a new fermion which has both Dirac and Majorana masses. 
The fermion $\chi$ initially has a Dirac mass $M_d$ and vector coupling with respect to the $Z'$. A Majorana mass can be generated from the vev of a $U(1)'$ Higgs through an interaction $y_\chi \Phi_\chi \bar\chi \chi^c$, so that
\begin{equation}
	{\cal L} \supset \bar \chi ( i \Dslash - M_d ) \chi  -  \frac{M_m}{2} (\bar \chi \chi^c + \text{h.c.}).
\end{equation}
This will lead to two Majorana states $\chi_{1,2}$ with masses $M_{1,2} = |M_m \pm M_d|$.  The interaction with the $Z'$ is off-diagonal and can be written as:
\begin{equation}
  \frac{g_\chi}{2} Z'_\mu  \left( \bar \chi_2 \gamma^\mu \gamma^5 \chi_1 +
     \bar \chi_1 \gamma^\mu \gamma^5 \chi_2 \right)
\end{equation}
As long as the splitting is large enough, it is possible to have the decay $\chi_2 \to Z' \chi_1$. For example, if the scalar giving rise to the Majorana mass is also the scalar responsible for $U(1)'$ breaking, $M_m$ can easily be of order $M_{Z'}$.  Here we have assumed a charge conjugation symmetry, such that there is only one Majorana mass; if there are different Majorana masses for left- and right-handed components, diagonal couplings are also present.

As in the previous model, we allow the $Z'$ couplings to quarks and leptons to be set by additional free parameters. Our assumption is that the $\chi_2$ has 100$\%$ branching to $\chi_1 Z'$, and that the $Z'$ has 100$\%$ branching to $q \bar q$, giving the final state signature of a dijet resonance plus missing transverse momentum. For the dilepton plus missing transverse momentum signature, we allow for a significant branching fraction of the $Z'$ to muons.
	
To avoid scanning over too many parameters, we consider two sets of  benchmarks for $M_{1,2}$. Since the cross section increases with lower $\chi_1$ mass, we include one optimistic case with very light $\chi_1$:
\begin{eqnarray}
	 M_1 = 5\ \GeV,  \ \ \  M_2 = M_1 + M_{Z'} + \Delta;\  \Delta = 25\ \GeV
	\label{eq:spectrum1}
\end{eqnarray}
This case is somewhat tuned for large $Z'$ mass, since it requires a cancellation between Dirac and Majorana masses. 

We also include a case where the fermion masses scale with $M_{Z'}$:
\begin{equation}
	M_1 = M_{Z'}/2, \ \ \ M_2 = 2 M_{Z'}
	\label{eq:spectrum2}
\end{equation}
With $M_1 < M_{Z'}$, the interactions above are not sufficient for $\chi_1$ to obtain the correct thermal abundance in the standard cosmology. Since this is model-dependent, we leave this an open question and instead focus here on lighter dark sector masses, where the LHC sensitivity is better.

\begin{figure*}
\vspace{2cm} \ \\
\includegraphics[width=0.47\linewidth]{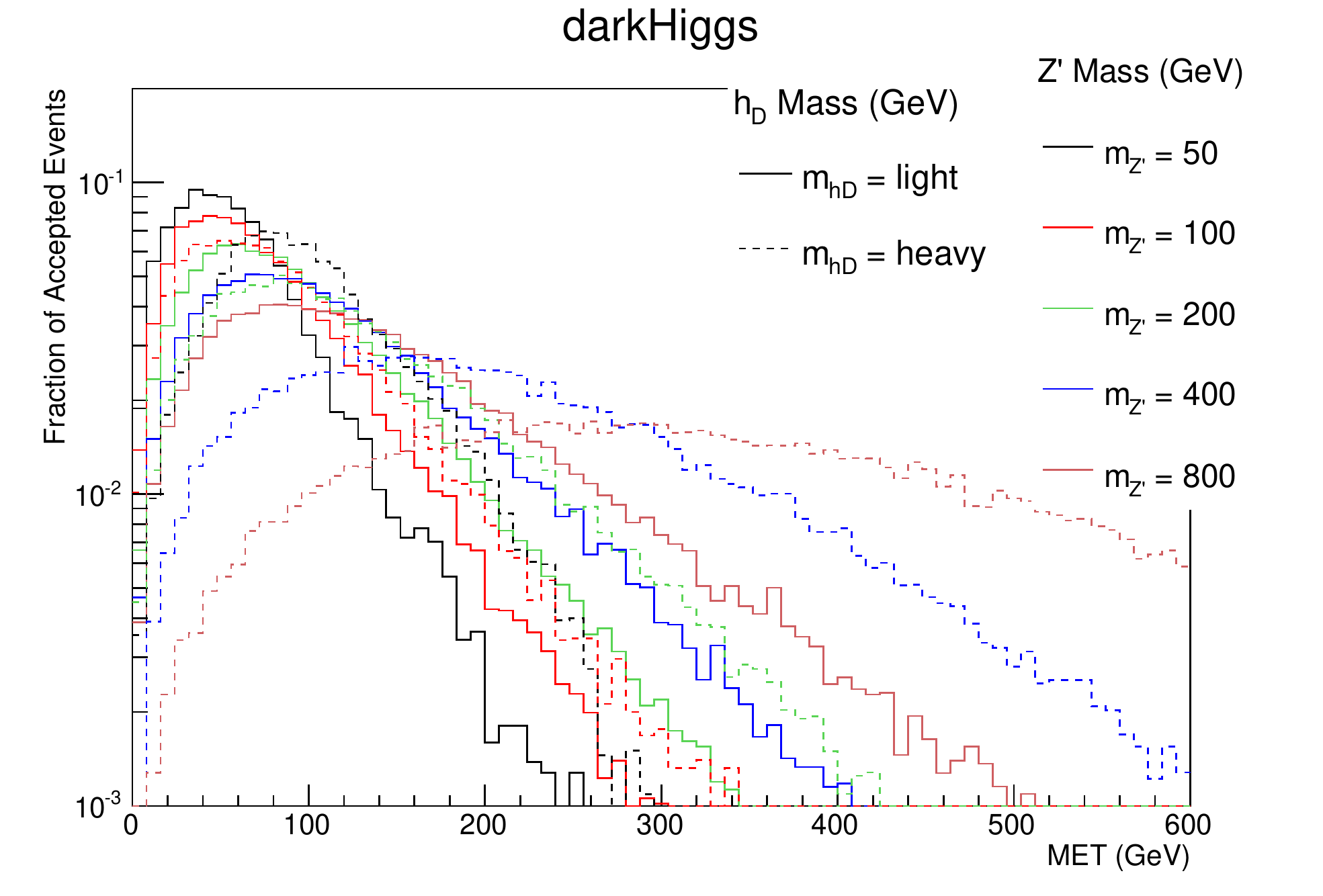}
\includegraphics[width=0.47\linewidth]{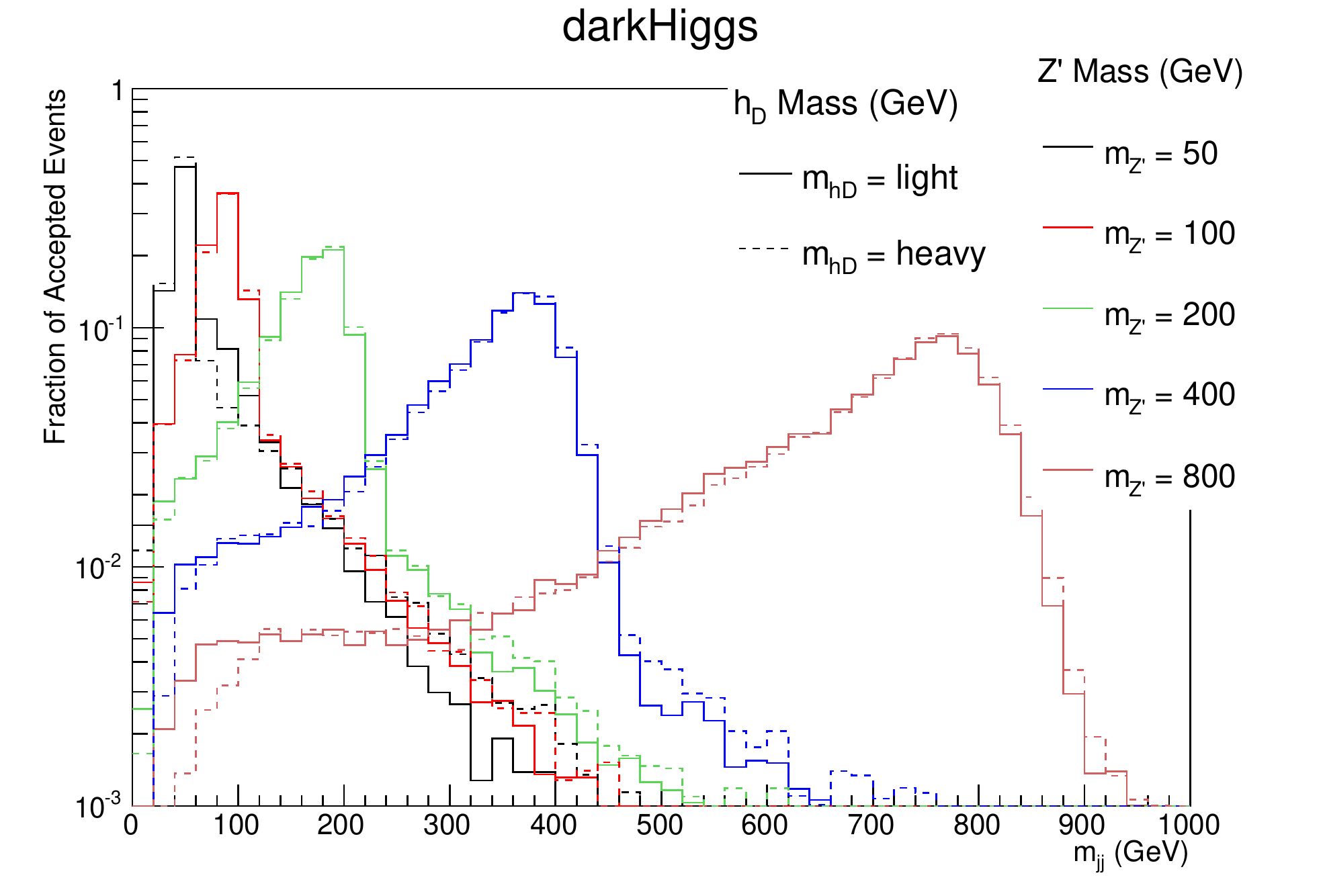}
\includegraphics[width=0.47\linewidth]{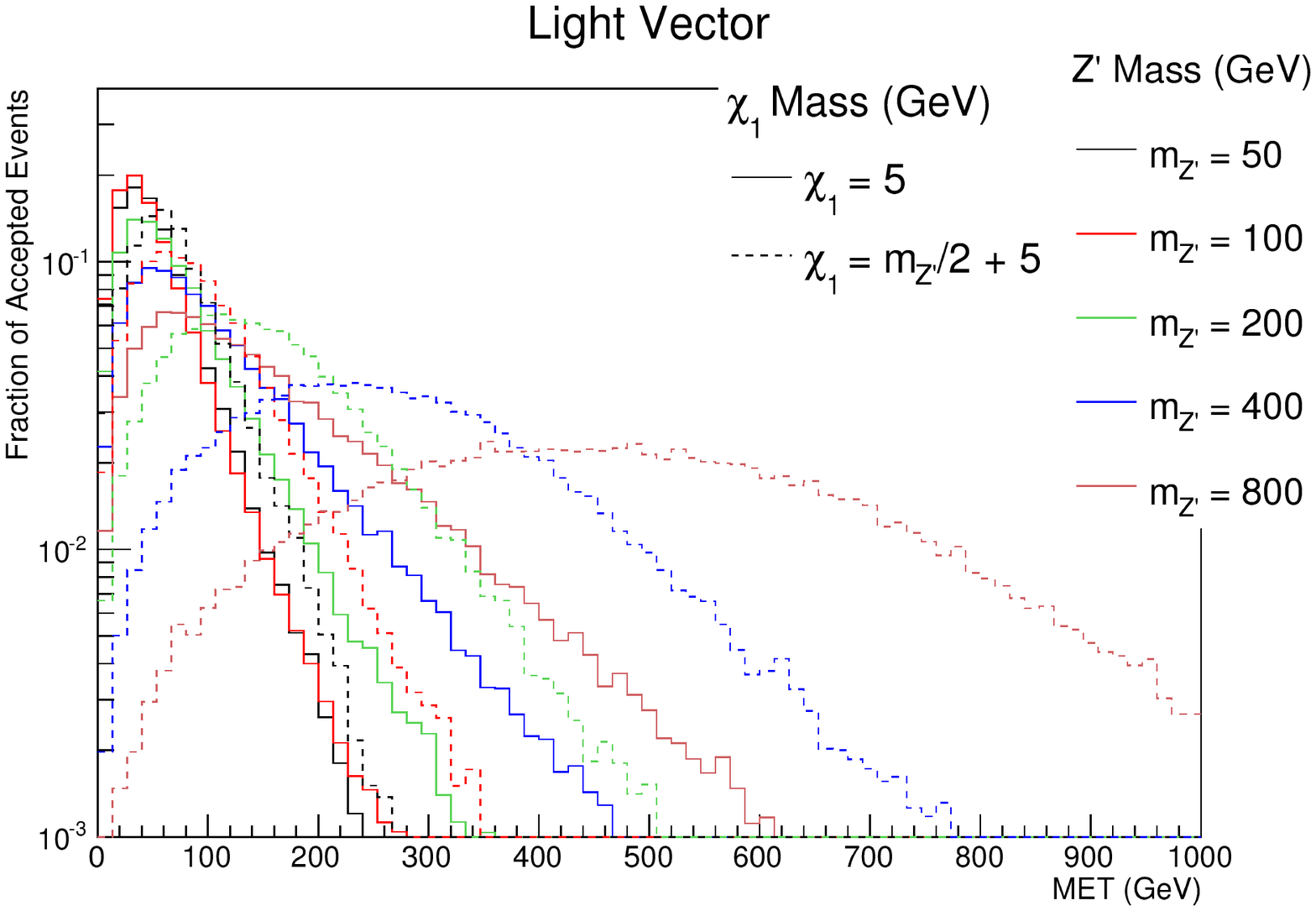}
\includegraphics[width=0.47\linewidth]{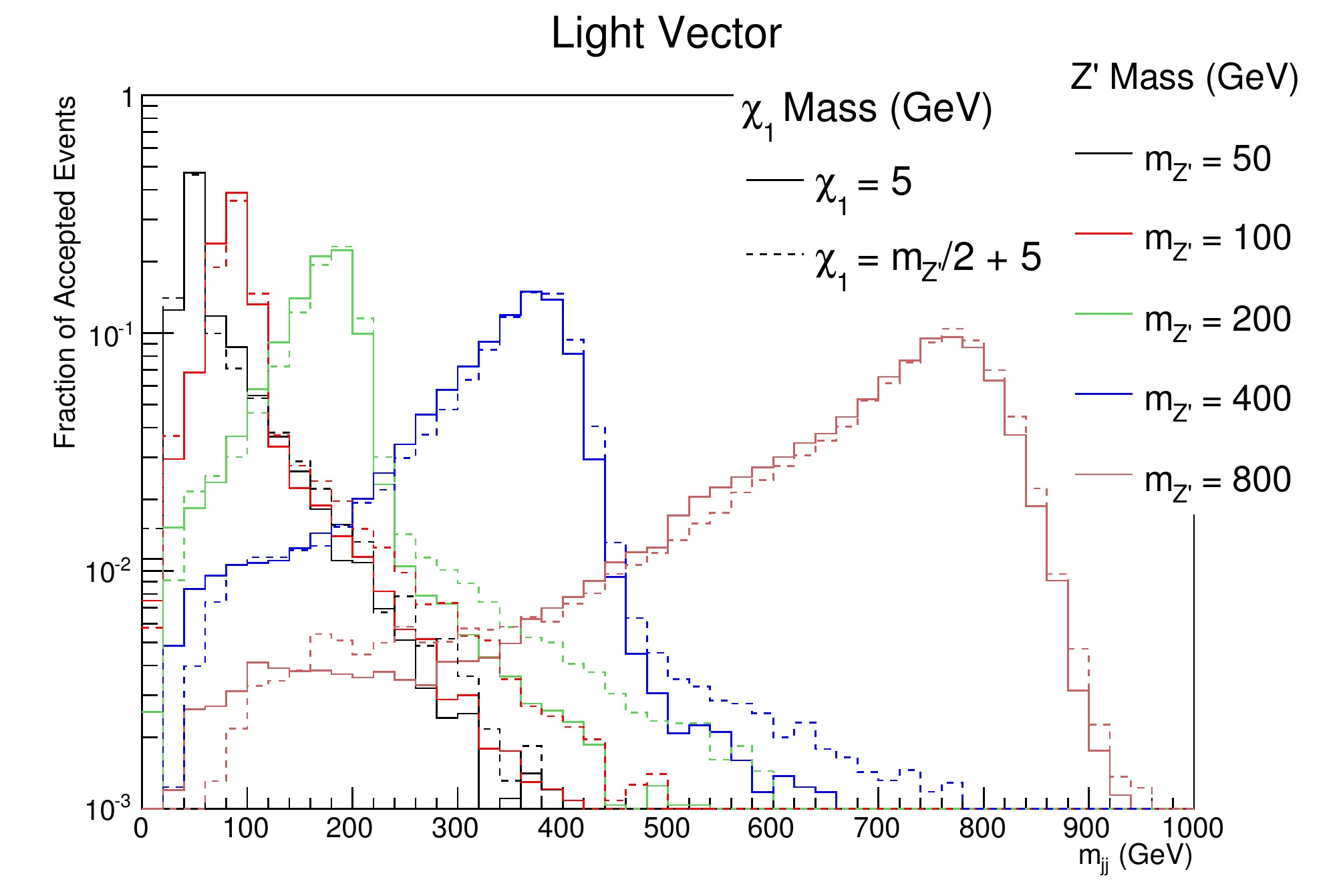}
\includegraphics[width=0.47\linewidth]{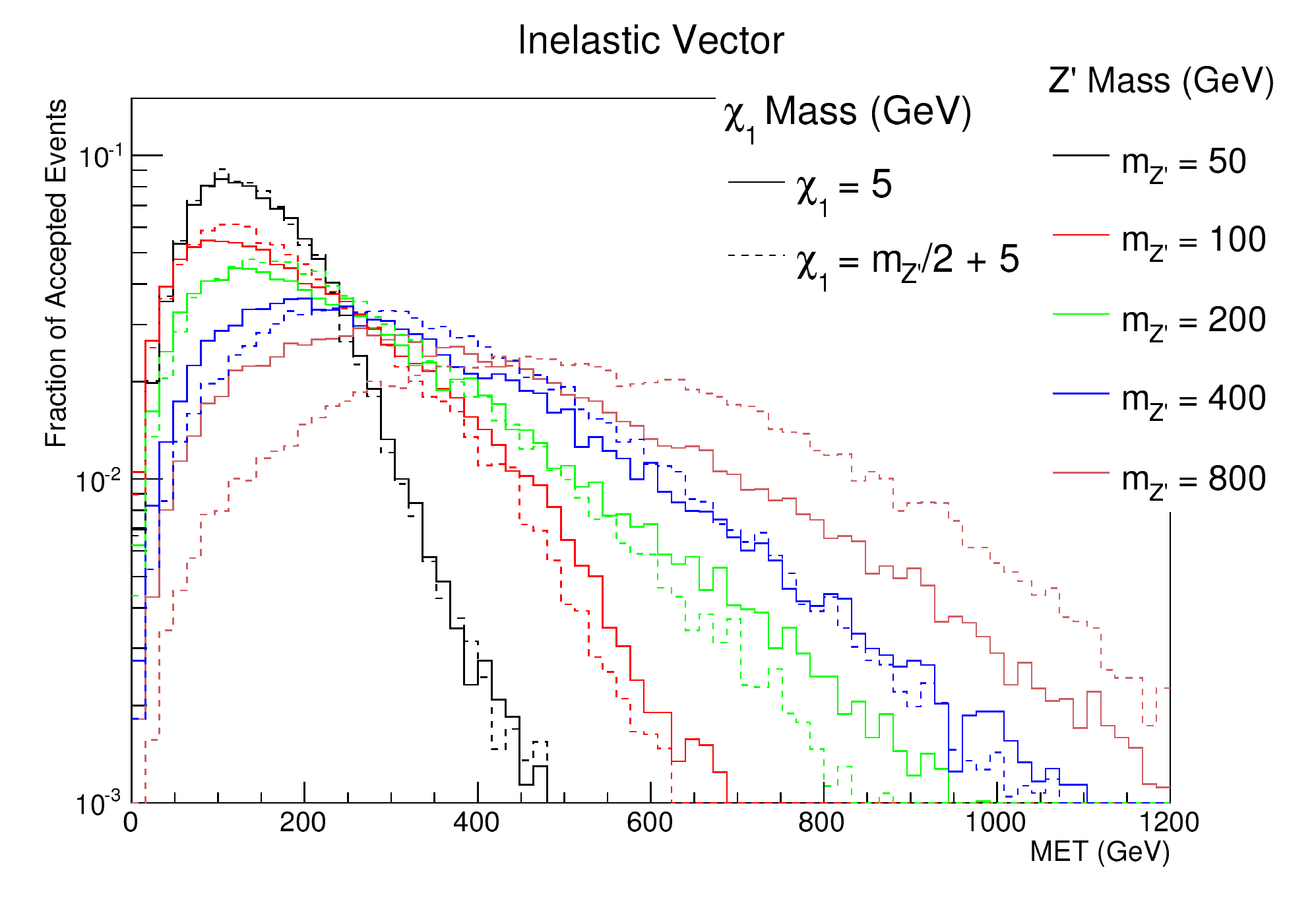}
\includegraphics[width=0.47\linewidth]{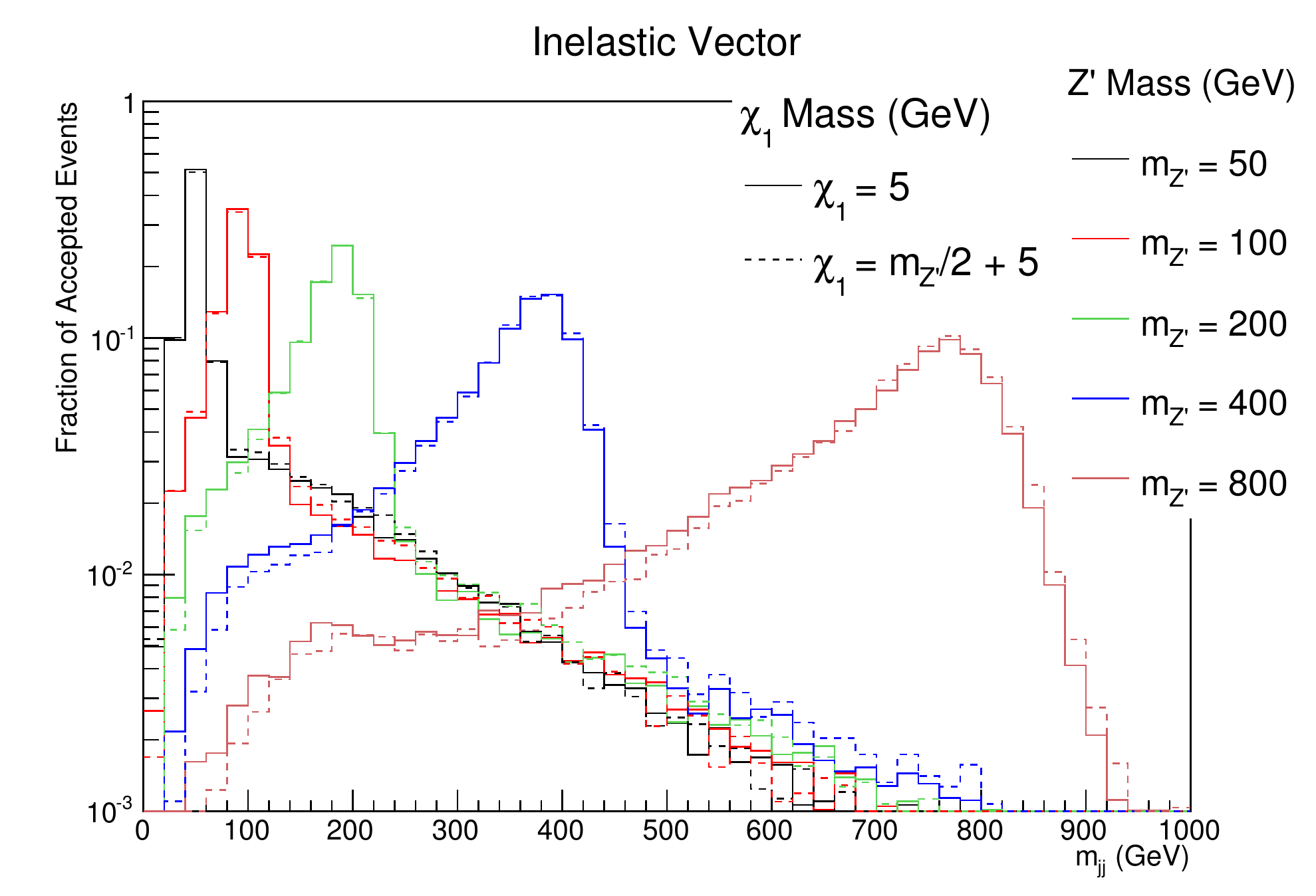}
\caption{ Distribution of reconstructed $\missET$ ({\it left})  and $m_{jj}$ ({\it right})   in the $jj+\missET$ final state for each of the three models considered. We show a subset of our $Z'$ mass points and consider the two cases for the masses of the other states, as discussed in the text.
\vspace{2cm}}
\label{fig:jj_kin}
\end{figure*}

\subsection{Light $Z'$ with Inelastic EFT coupling}

The models thus far rely on the $Z'$ coupling to quarks in order to be produced at the LHC. Rather than producing dark sector states through the new $Z'$, we consider the possibility that it is produced through a new contact interaction:
\begin{equation}
	\frac{1}{2 \Lambda^2} \bar q \gamma^\mu q  \left( \bar \chi_2 \gamma^\mu \gamma^5 \chi_1 +
     \bar \chi_1 \gamma^\mu \gamma^5 \chi_2 \right).
	\label{eq:inelasticvector}
\end{equation}
Similar to the model just discussed, we have assumed two dark sector states $\chi_{1,2}$ with an off-diagonal coupling to the $Z'$. The $Z' + \missET$ process is analogous to that of the previous section; however, we have effectively replaced the intermediate $s$-channel $Z'$ with a heavy $Z_H'$, where the $Z_H'$ has been integrated out to give the operator above. For our benchmarks, the mass spectrum of the states is taken be the same as in Eqs.~\ref{eq:spectrum1} and \ref{eq:spectrum2}.

The $Z'$ produced in the decay can then be very weakly coupled to SM fermions, evading many direct search constraints. 
For example, this small coupling could be generated by kinetic mixing of the $Z'$ with hypercharge and kinetic mixing parameter $\epsilon \ll 1$. 
The only requirement is that the $Z'$ decays to the visible fermions on collider time scales, which is easily satisfied for $\epsilon \gtrsim 10^{-5}$. For each search channel we show results assuming either a 100$\%$ branching fraction to $jj$ or $\mu\mu$ in order to match our signal regions; however the results can easily be scaled for the case of kinetic mixing where, for example, ${\rm Br}(\mu\mu) \approx 0.12$ for large $M_{Z'}$.

Similar ideas have been considered in hidden valley models~\cite{Strassler:2006im,Han:2007ae}, which can give lepton jet signals from multiple light $Z'$s~\cite{Bai:2009it,Chan:2011aa}. The main difference here is a looser signal requirement of a single $Z'$ in the final state, and a wider range of $Z'$ masses considered, which allow reconstruction of the dijet or dilepton resonance.

\section{LHC Sensitivity}

In the following sections, we consider the $Z'\rightarrow jj$ and $Z'\rightarrow \ell\ell$ decay modes, propose an event selection and describe the expected sensitivity of the LHC dataset to $Z' + \missET$ for each of the models above.

\subsection{Dijet Mode}

Decays of a $Z'$ to a pair of quarks results in two high-$p_{\textrm{T}}$ jets.  In the following, the basic preselection requires at least two jets, each with $p_{\textrm{T}}>20$ GeV and $|\eta|<2.5$. Events with a reconstructed electron or muon with $p_{\textrm{T}}>10$ and $|\eta|<2.5$ are vetoed.

The candidate $Z'$ is reconstructed from the leading two $p_{\textrm{T}}$ jets. To suppress the non-peaking backgrounds, a mass window $m_{jj} \in [0.8 \times m_{Z'},m_{Z'}+30\ \textrm{GeV}]$ is applied.  Distributions of $m_{jj}$ and $\missET$ for the signal are shown in Fig.~\ref{fig:jj_kin}. For further details on how these distributions vary among the models, see the Discussion section.

\begin{figure}[t!]
\includegraphics[width=\linewidth]{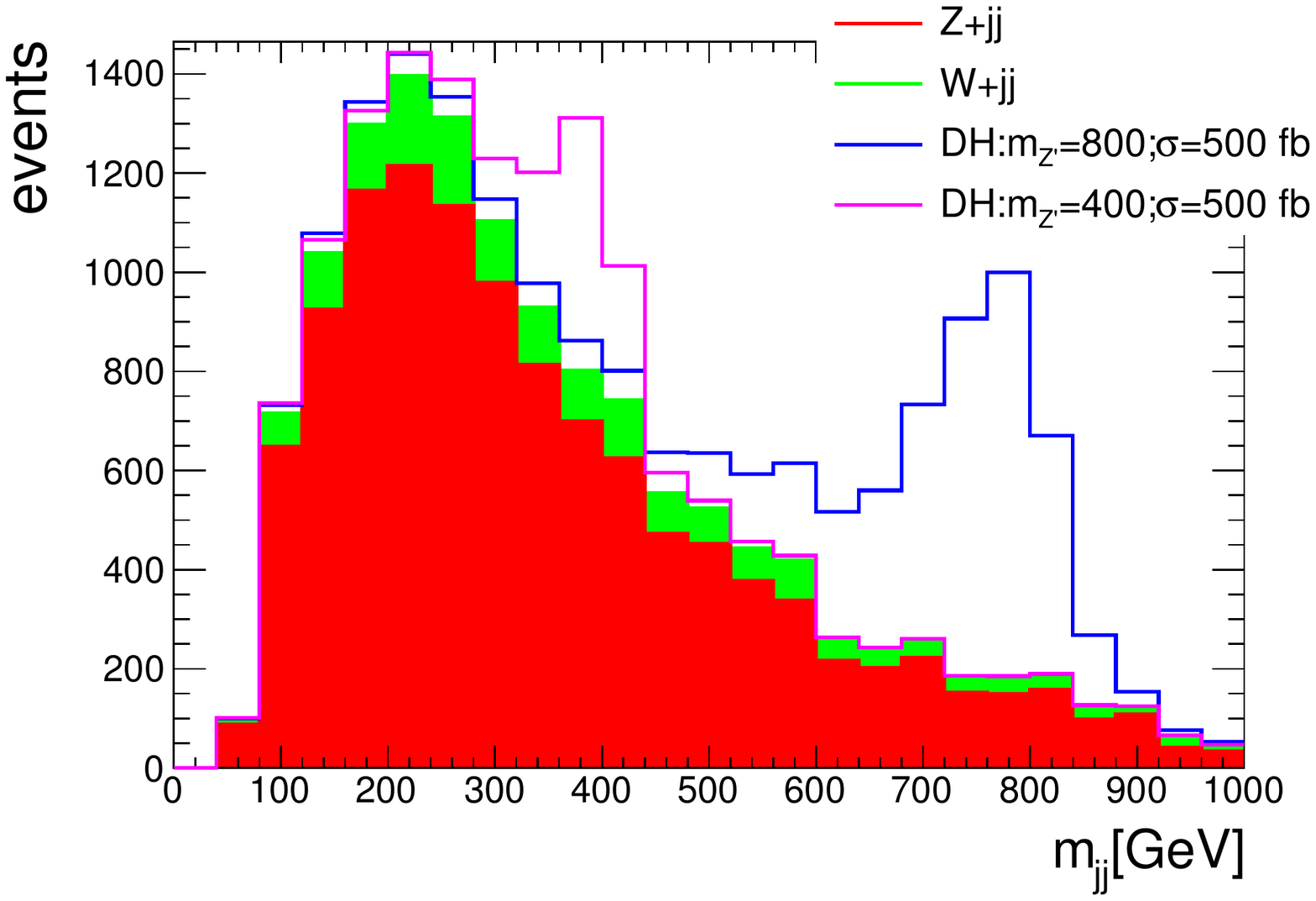}
\caption{ Distribution of reconstructed $m_{jj}$ in the $jj+\missET$ final state, for the expected SM background as well as several examples of the signal in the dark Higgs (DH) model. Events are required to satisfy the preselection as well as have $\missET>300$ GeV and leading jet $p_{\textrm{T}}>250$ GeV, but no $m_{jj}$ selection is applied.}
\label{fig:jj_bg}
\end{figure}

The primary background processes are $Z\rightarrow \nu\nu$ in association with two initial-state jets, or $W\rightarrow\ell\nu$ in association with two initial-state jets and where the charged lepton is not identified.  Events are simulated at parton level with {\sc madgraph}5~\cite{Alwall:2014hca}, with {\sc pythia}~\cite{Sjostrand:2006za} for showering and hadronization and {\sc delphes}~\cite{deFavereau:2013fsa} with the ATLAS-style configuration for detector simulation.  Backgrounds are normalized to leading-order cross sections; the uncertainty is calculated by varying the factorization and renormalization scales by factors of two.  We validate our background model by comparing to the ATLAS results~\cite{Aad:2013oja} with $m_{jj} \in [50,120]$ GeV and $\missET>$ 350,500 GeV; the comparison is not precise due to the differences in the jet algorithm and radius parameters, but the estimates are roughly consistent.  In Fig.~\ref{fig:jj_bg}, distributions of $m_{jj}$ are shown for the expected backgrounds.  

\begin{figure}[t!]
\includegraphics[width=1.0\linewidth]{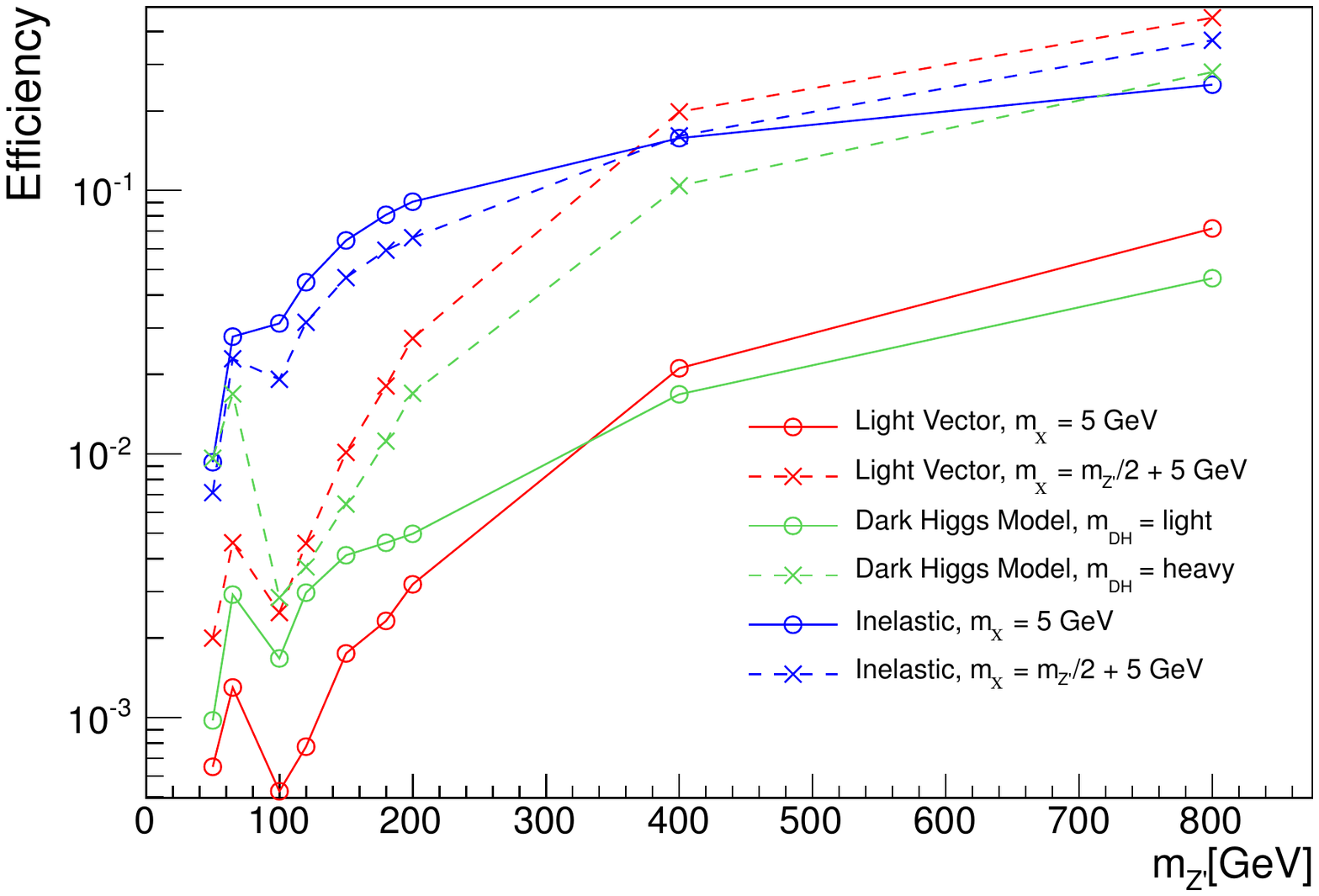}
\includegraphics[width=0.97\linewidth]{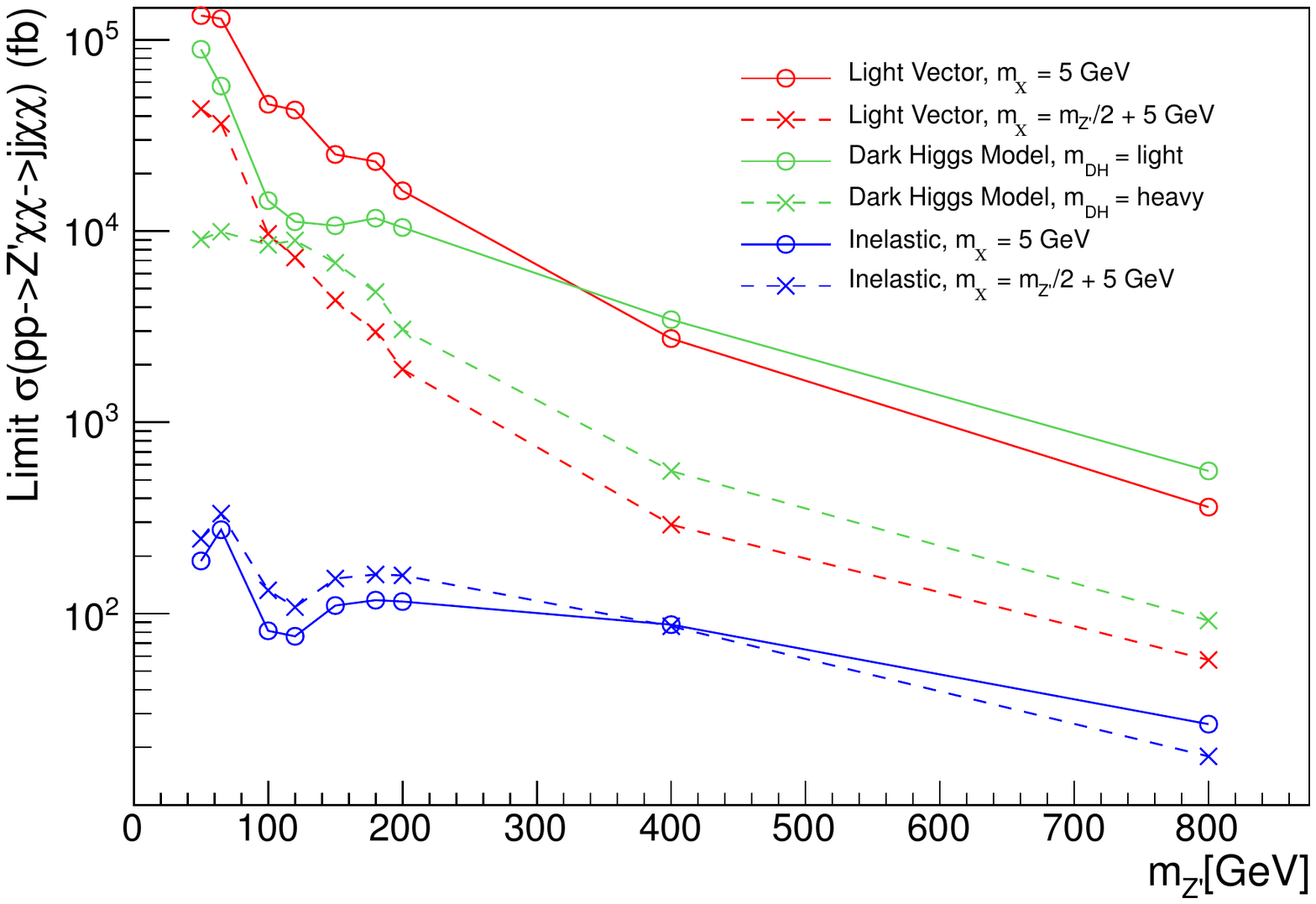}
\caption{ ({\it Top}) Efficiency of the $jj+\missET$ selection described in the text, for two choices of mass spectra in each of the three models considered. Note that the minimum required $\missET$ increases above $m_{Z'}>100$ GeV. ({\it Bottom}) 95$\%$ CL upper limits on the production of $Z'\rightarrow jj+\missET$ as a function of the $Z'$ mass.}
\label{fig:jj_limits}
\end{figure}

To suppress the large dijet background, large $\missET$ is required.  The value of the threshold in $\missET$ is determined by optimizing with respect to the expected upper limits on the cross section.  In the case of the dark Higgs and light vector models, which have similar $\missET$ distributions, the threshold is $\missET> 200 (300)$ GeV for values of $m_{Z'} < 100 $  ($ >100$) GeV. In the case of the inelastic EFT model, which has larger  $\missET$, the threshold is $\missET> 300 (400)$ GeV for values of $m_{Z'} < 100$ ($>100$) GeV.  In addition, we require the $p_{\textrm{T}}$ of the leading jet to be at least $(\missET^{\textrm{thresh}} - 50)$ GeV, which helps in suppressing the $V$+jets background.  The efficiency of the final selection is shown in Fig~\ref{fig:jj_limits} and detailed in Table~\ref{tab:jj_estimates} for various $Z'$ and dark matter masses.

Upper limits are calculated in counting experiments, using a profile likelihood ratio~\cite{Cowan:2010js} with the CLs technique~\cite{Read:2002hq,Junk:1999kv}.  Limits on the production cross section of $\sigma(pp\rightarrow Z'\chi\bar{\chi}\rightarrow jj\chi\bar{\chi})$ are shown in Fig.~\ref{fig:jj_limits}.

\begin{table}[t!]
\caption{ Signal efficiency and expected background yields for several $Z'$ masses in the $jj+\missET$ final state. Only the heavy mass spectrum choice is listed. The background uncertainty is 27\% obtained by varying the renormalization and factorization scales by factors of two. }
\center
\begin{tabular}{l|rrrrrr}
\hline \hline
& \multicolumn{3}{c}{$m_{Z'}$ [GeV]} \\ 
& 50 & 200 & 400 \\ \hline\hline
$\missET$ [GeV] & $>200$ & $>300$ & $>300$ \\ \hline

& \multicolumn{3}{c}{Signal Efficiencies} \\
Dark Higgs & 0.01 & 0.02 & 0.10\\
Light Vector & 0.002 & 0.03 & 0.20\\
\hline
& \multicolumn{3}{c}{Background Estimates} \\
$Z\rightarrow \nu\nu +jj$ & 3000 & 2,200& 2,000 \\
$W\rightarrow \ell\nu +jj$ & 350 &  300& 330\\
Total Background & 3,350 & 2,500 & 2,300 \\
\hline\hline
$\missET$ [GeV] & $>300$ & $>400$ & $>400$ \\ \hline
& \multicolumn{3}{c}{Signal Efficiencies} \\ 
Inelastic EFT& 0.007 & 0.07 & 0.16\\\hline
& \multicolumn{3}{c}{Background Estimates} \\
$Z\rightarrow \nu\nu +jj$ &  60 & 360 & 470 \\
$W\rightarrow \ell\nu +jj$ &  10 &  50 & 65 \\
Total Background & 70 & 410 & 535 \\
\hline \hline
\end{tabular}
\label{tab:jj_estimates}
\end{table}

\begin{figure*}
\vspace{2cm} \ \\
\includegraphics[width=0.45\linewidth]{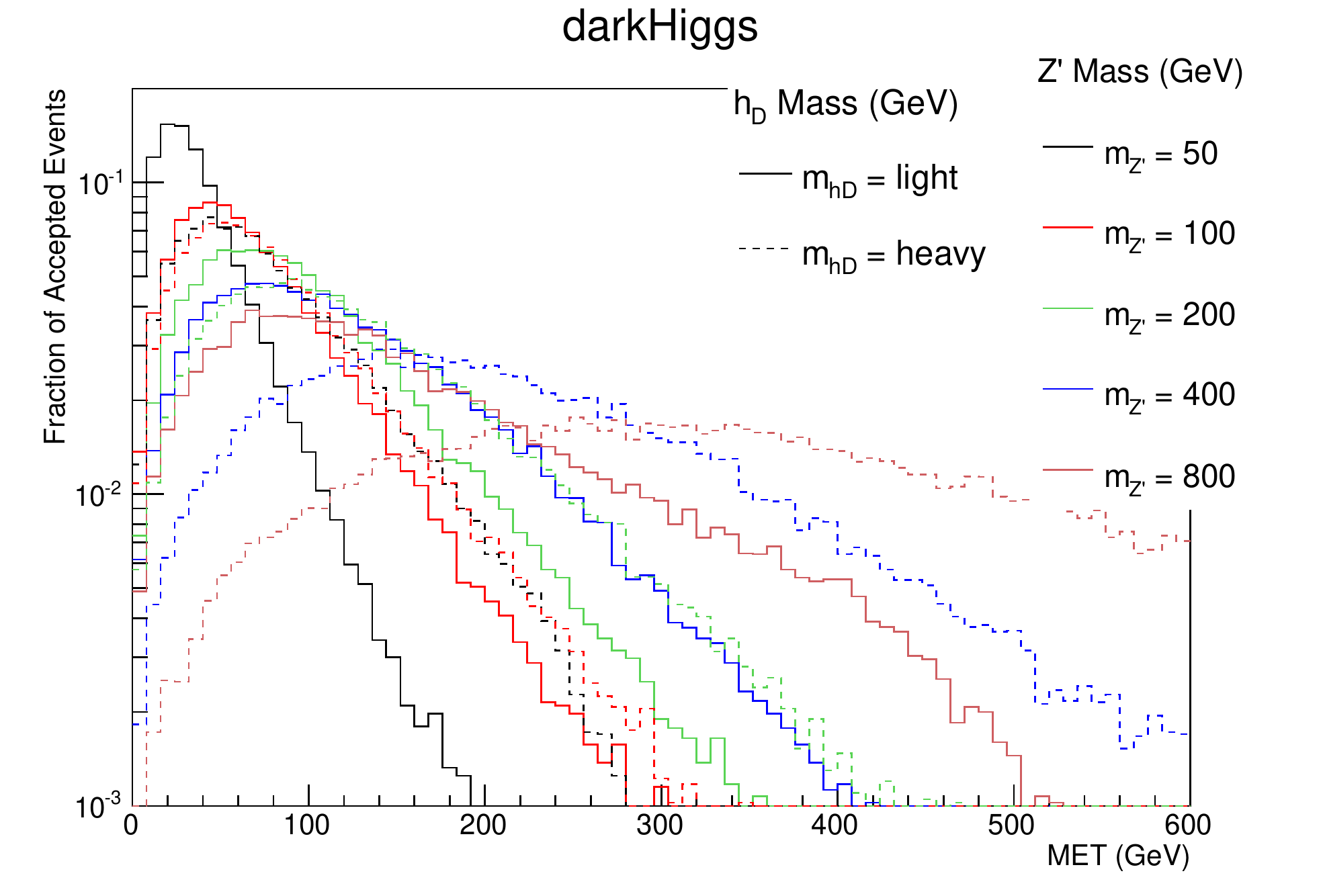}
\includegraphics[width=0.45\linewidth]{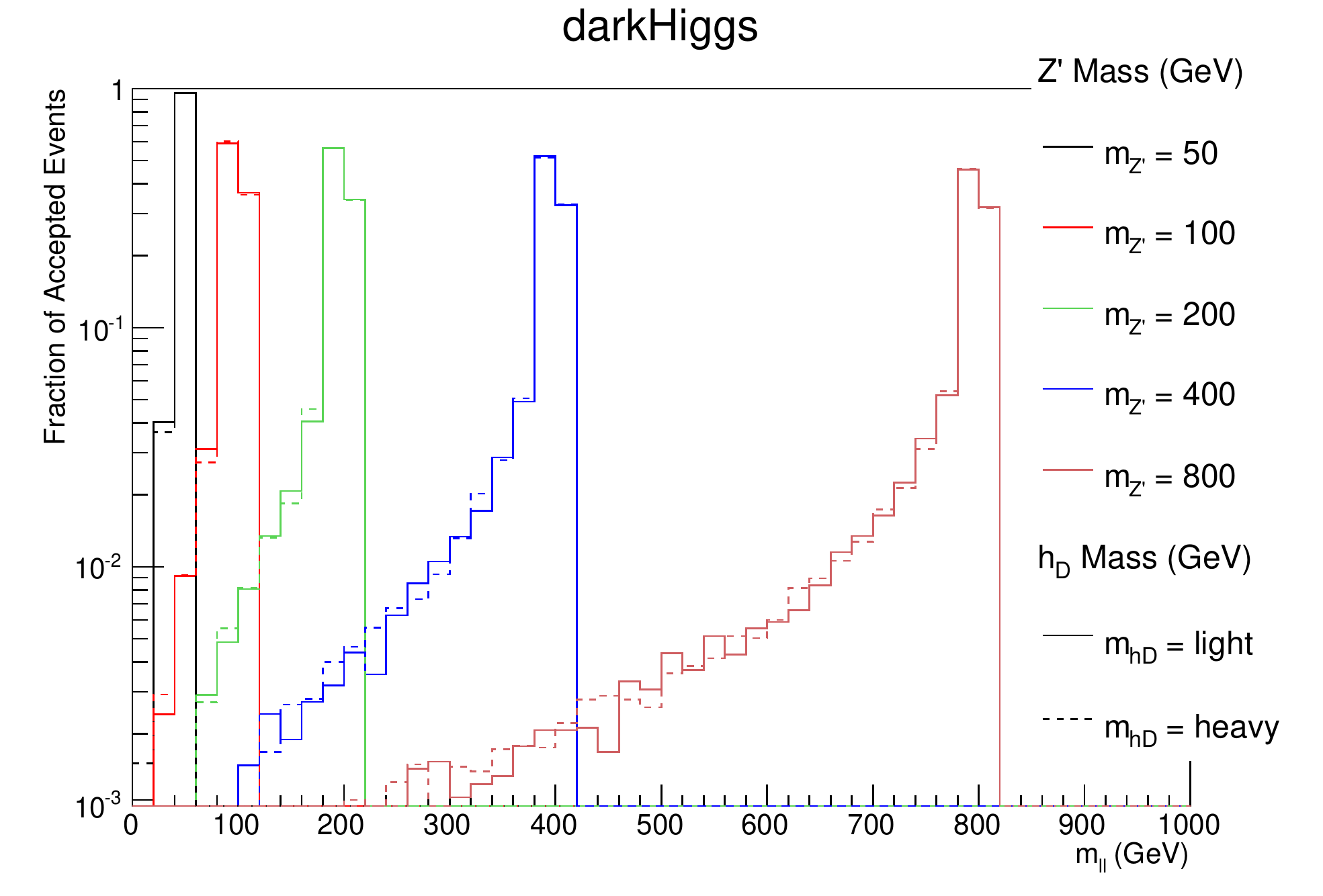}
\includegraphics[width=0.45\linewidth]{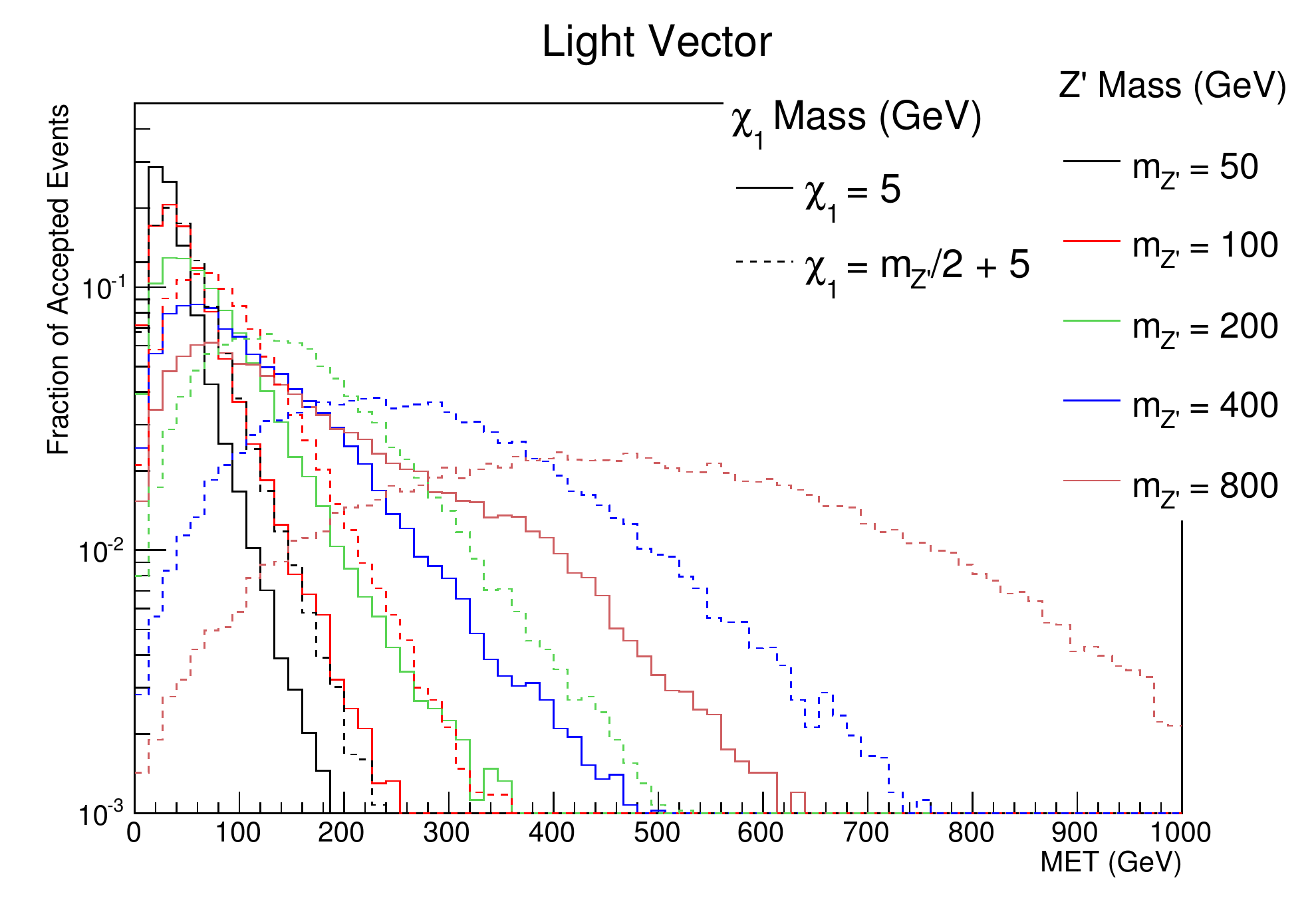}
\includegraphics[width=0.45\linewidth]{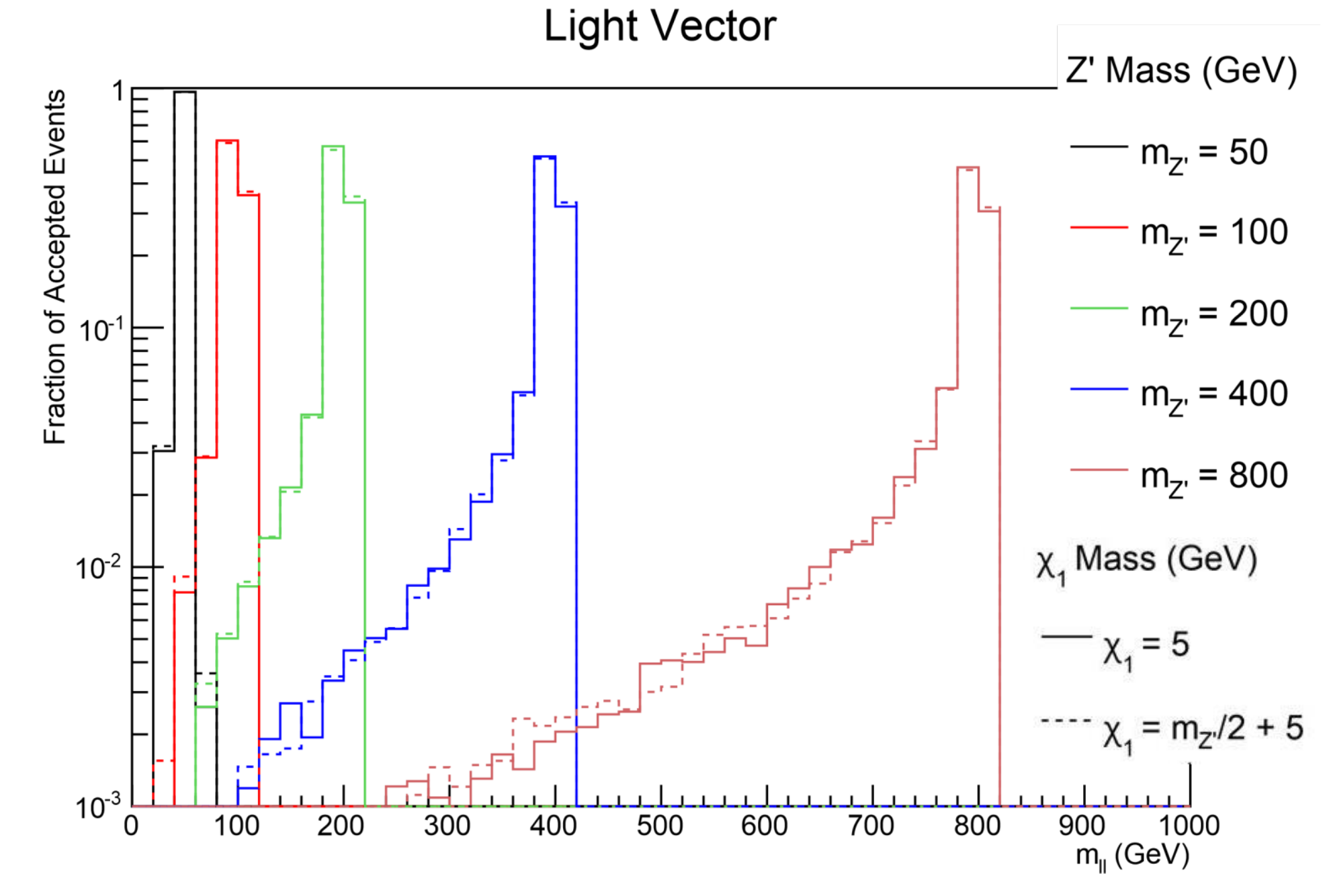}
\includegraphics[width=0.45\linewidth]{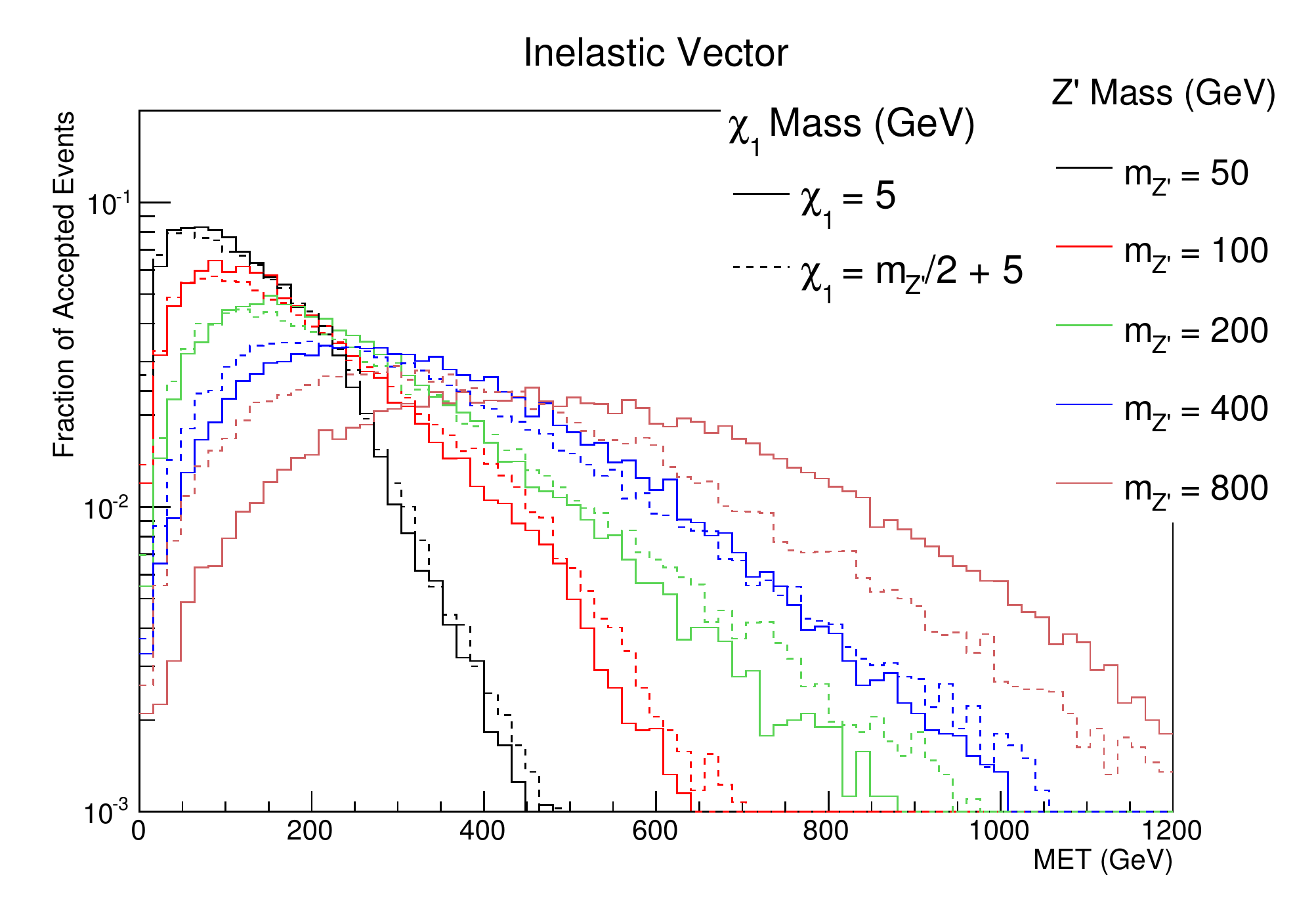}
\includegraphics[width=0.45\linewidth]{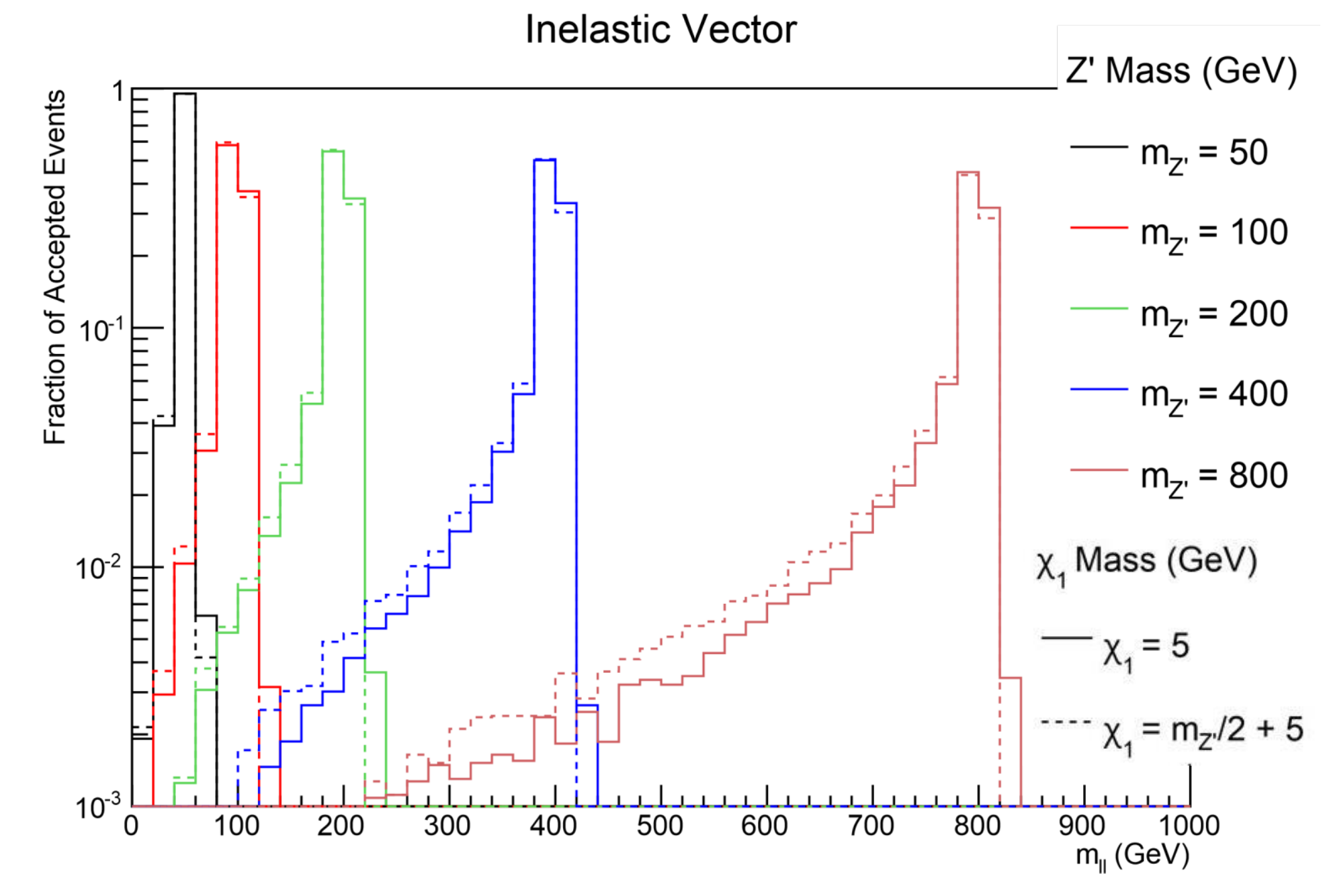}
\caption{ Distribution of reconstructed $\missET$ ({\it left}) and  $m_{\ell\ell}$ ({\it right}) in the $\ell\ell+\missET$ final state for each of the three models considered. We show a subset of our $Z'$ mass points and consider the two cases for the masses of the other states, as discussed in the text.
\vspace{2cm}}
\label{fig:ll_kin}
\end{figure*}

\subsection{Dilepton mode}

Leptonic decays of a $Z'$ may result in two high-$p_{\textrm{T}}$ electrons or muons.  In the following, the basic preselection requires at least two opposite-sign electrons or muons, each with $p_{\textrm{T}}>30$ GeV and $|\eta|<2.5$ as well as $\missET>100$ GeV and $p_{\textrm{T}}(\ell \ell) > 80\ \textrm{GeV}$.  Events with a third charged lepton or at least one jet with $p_{\textrm{T}}>20$ GeV and $|\eta|<2.5$ are vetoed.  

Due to the tight constraints on $Z'$ coupling to electrons discussed above, we will focus on the muonic channel here. To a good approximation, the backgrounds would be larger by a factor of 2 if both lepton final states were included, and for models where the $Z'$ decays to both charged lepton flavors, the resulting limits would be stronger by up to a factor of  $\sqrt{2}$ if systematic uncertainties are not dominant.

The candidate $Z'$ is reconstructed from the two leptons.   To suppress backgrounds which do not peak at the $Z'$ mass,  a requirement that $m_{\ell\ell} \in [0.9 \times m_{Z'},m_{Z'}+25\ \textrm{GeV}]$ is applied. Distributions of $m_{\ell\ell}$ and $\missET$ are shown in Fig.~\ref{fig:ll_kin}; the dependence of these on different models and mass parameter choices is examined further in the Discussion section.

\begin{figure}[t!]
\includegraphics[width=\linewidth]{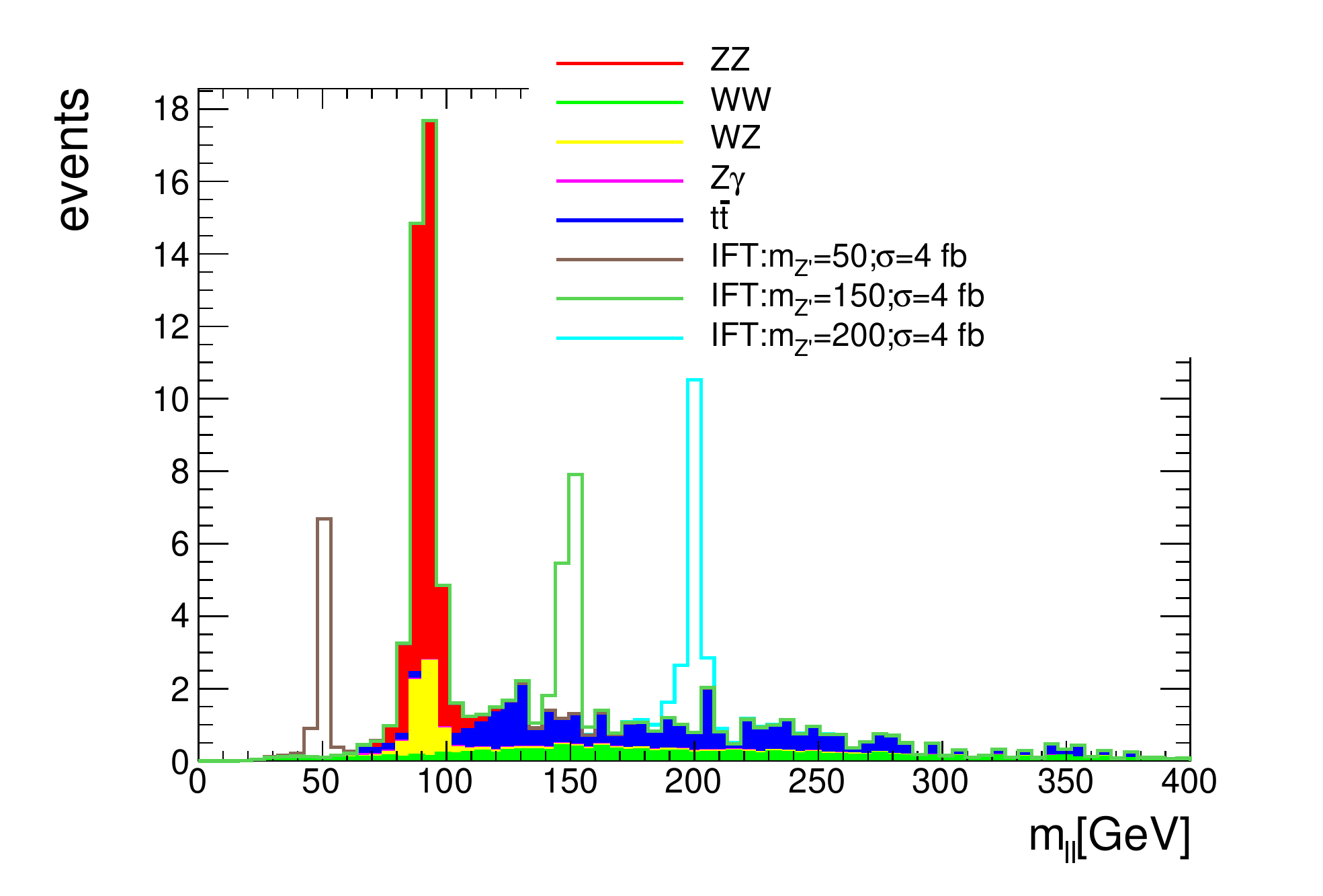}
\caption{ Distribution of reconstructed  $m_{\ell\ell}$ in the $\mu^+\mu^-+\missET$ final state, for the expected background as well as a signal example.  The IFT label refers to the inelastic EFT model.  Events are required to satisfy the preselection as well as have $\missET>100$ GeV and $p_{\ell\ell}>80$ GeV, but no $m_{\ell\ell}$ selection is applied.}
\label{fig:ll_bg}
\end{figure}

The primary background processes are diboson production, such as $ZZ\rightarrow \ell\ell\nu\nu$, $WZ\rightarrow\ell\nu\ell\ell$, $WW\rightarrow\ell\nu\ell\nu$ or $Z\gamma\rightarrow \ell\ell\nu\nu$.  Top pair backgrounds are effectively suppressed via the jet veto. Events are simulated at parton level with {\sc madgraph}5~\cite{Alwall:2014hca}, with {\sc pythia}~\cite{Sjostrand:2006za} for showering and hadronization and {\sc delphes}~\cite{deFavereau:2013fsa} for detector simulation.  Backgrounds are normalized to leading-order cross sections; the uncertainty is calculated by varying the factorization and renormalization scales by factors of two. A minimum 15\% uncertainty is applied to cover uncertainty due to the high-$p_{\textrm{T}}$ region. We validate our background model by comparing to the ATLAS results~\cite{Aad:2014vka} with $m_{ll} \in [76,106]$ GeV and $\missET>$ 150, 250, 350, 450 GeV; our estimates agree within uncertainties.  In Fig.~\ref{fig:ll_bg}, distributions of $m_{\ell\ell}$ are shown with the expected background.

\begin{figure}[t!]
\includegraphics[width=\linewidth]{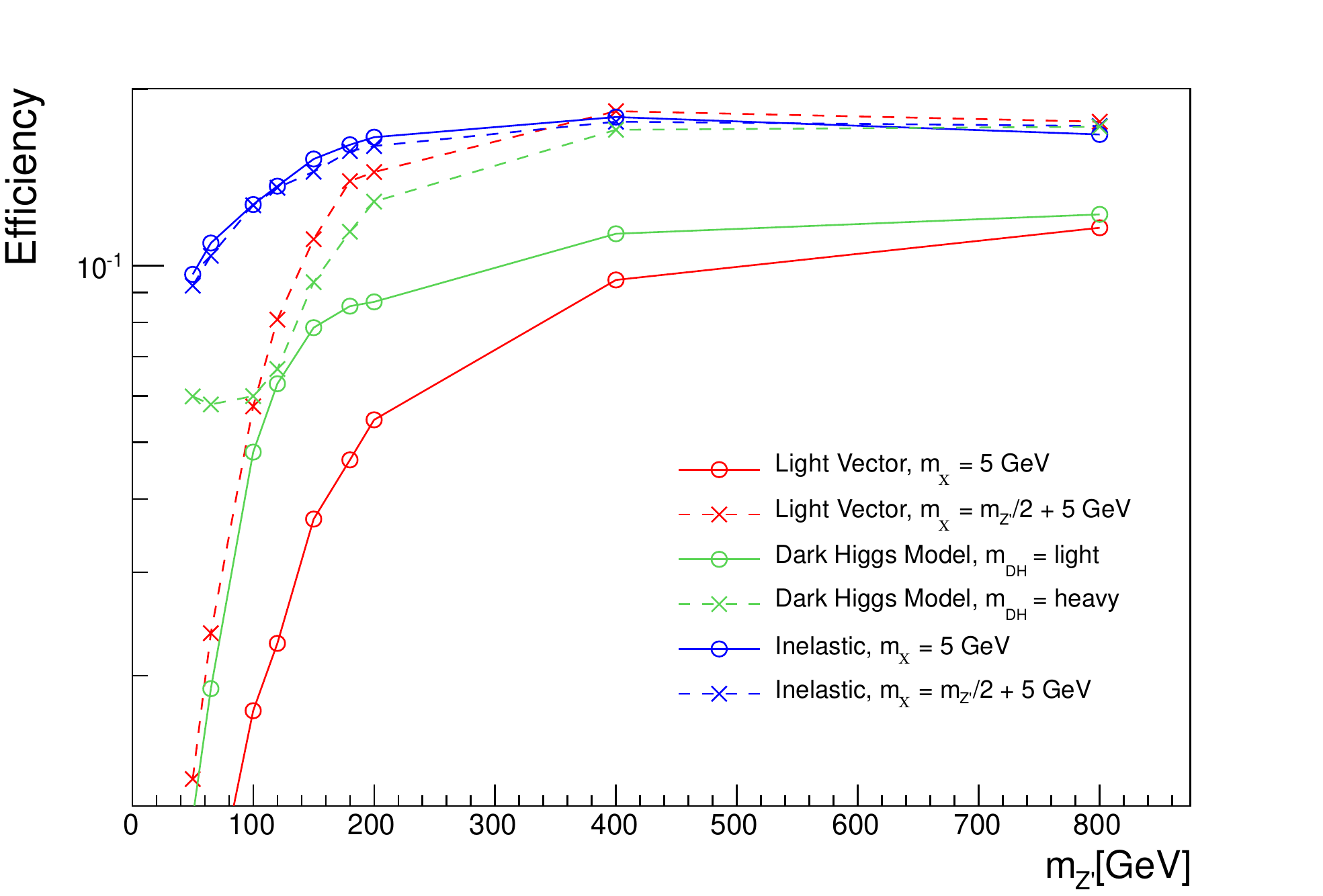}\\
\includegraphics[width=\linewidth]{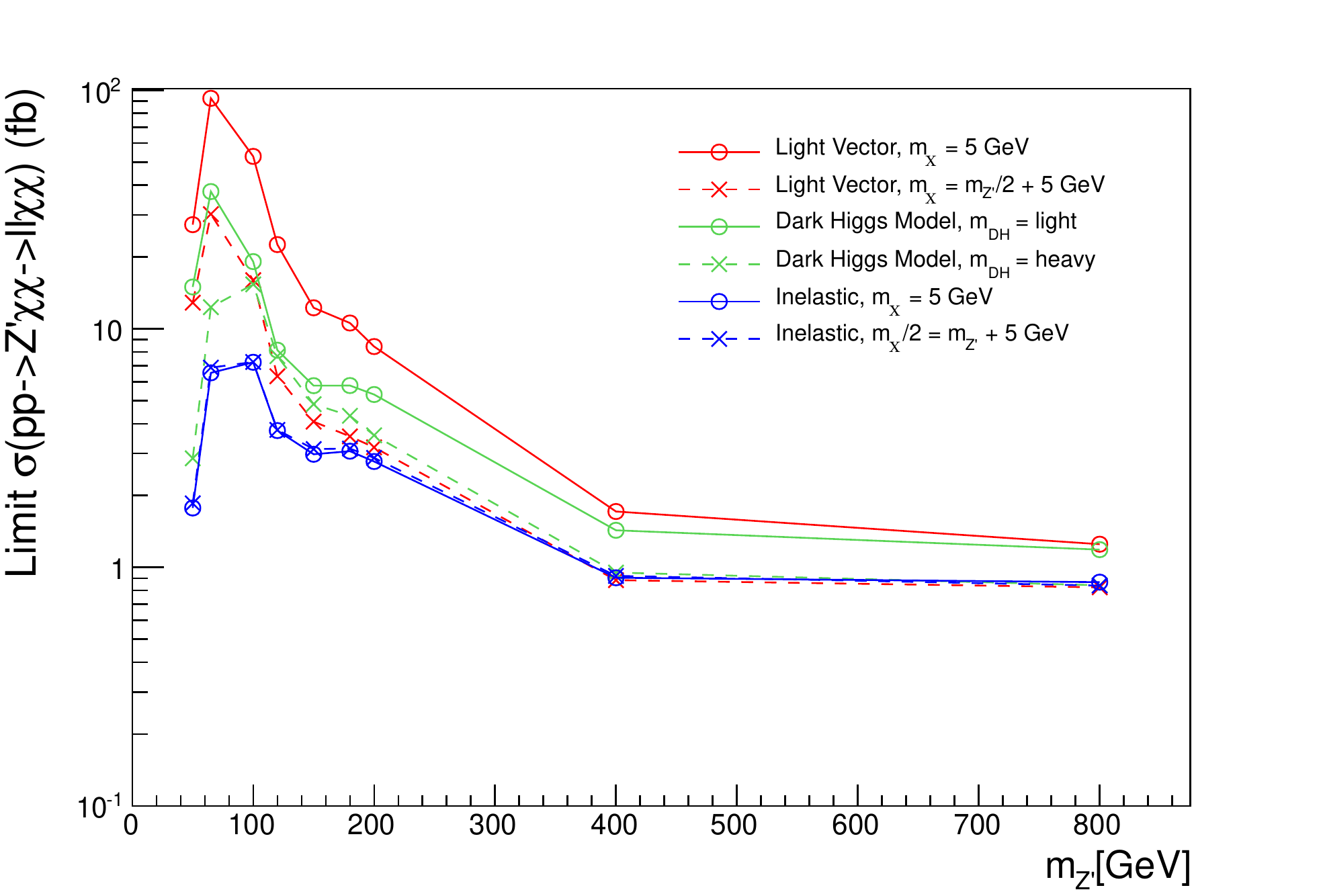}
\caption{ ({\it Top}) Efficiency of the $\mu^+\mu^-+\missET$ selection described in the text as a function of the $Z'$ mass, for two choices of mass spectra in each of the three models considered.  ({\it Bottom}) 95$\%$ CL upper limits on the production of $(Z'\rightarrow \mu^+\mu^-)+\missET$ as a function of the $Z'$ mass.}
\label{fig:ll_limits}
\end{figure}

As in the dijet case, large missing transverse momentum is required to suppress the large $\ell\ell$ backgrounds; the requirement $\missET>100$ is found to give the strongest expected limits across all models and masses.  The efficiency of the selection is shown in Fig~\ref{fig:ll_limits} and detailed in Table~\ref{tab:ll_estimates} for various $Z'$ and dark matter masses. 

Upper limits are calculated in counting experiments, using a profile likelihood ratio~\cite{Cowan:2010js} with the CLs technique~\cite{Read:2002hq,Junk:1999kv}.  Limits on the production cross section of $\sigma(pp\rightarrow Z'\chi\bar{\chi}\rightarrow \mu^+\mu^- \chi\bar{\chi})$ are shown in Fig.~\ref{fig:ll_limits}.

\begin{table}
\caption{ Signal efficiency and expected background yields for several $Z'$ masses in the $\mu^+\mu^-+\missET$ final state with $\missET>100$ GeV. In each model, the masses are are chosen to be that of the heavy spectrum case.}
\center
\begin{tabular}{l|C{1cm}C{1cm}R{1cm}}
\hline \hline
& \multicolumn{3}{c}{$m_{Z'}$ [GeV]} \\
			& 50 & 200 & 400 \\ \hline\hline
Model & \multicolumn{3}{c}{Signal Efficiencies} \\ \hline
Dark Higgs 	& 0.06 & 0.13 & 0.17 \\
Light Vector 	& 0.01 & 0.14 & 0.18\\
Inelastic EFT	& 0.09 & 0.16 & 0.18 \\
\hline\hline
Process & \multicolumn{3}{c}{Background Estimates} \\ \hline
$ZZ$   		  & 0.4 & -- & --\\
$WZ$  		  & 0.1 & 0.3 & 0.1\\
$WW$ 		  & 0.4 & 2.1 & 0.9\\
$Z\gamma^*$ & 0.3 & 0.1 & -- \\
$t\bar t$ 		  & 0.3 & 6.1 & 0.3\\\hline
Total Background & 1.6 & 8.6 & 1.3\\
\hline \hline
\end{tabular}
\label{tab:ll_estimates}
\end{table}

\section{Discussion \label{sec:disc}}

\begin{figure*}[t]
\centering
\includegraphics[width=0.47\linewidth]{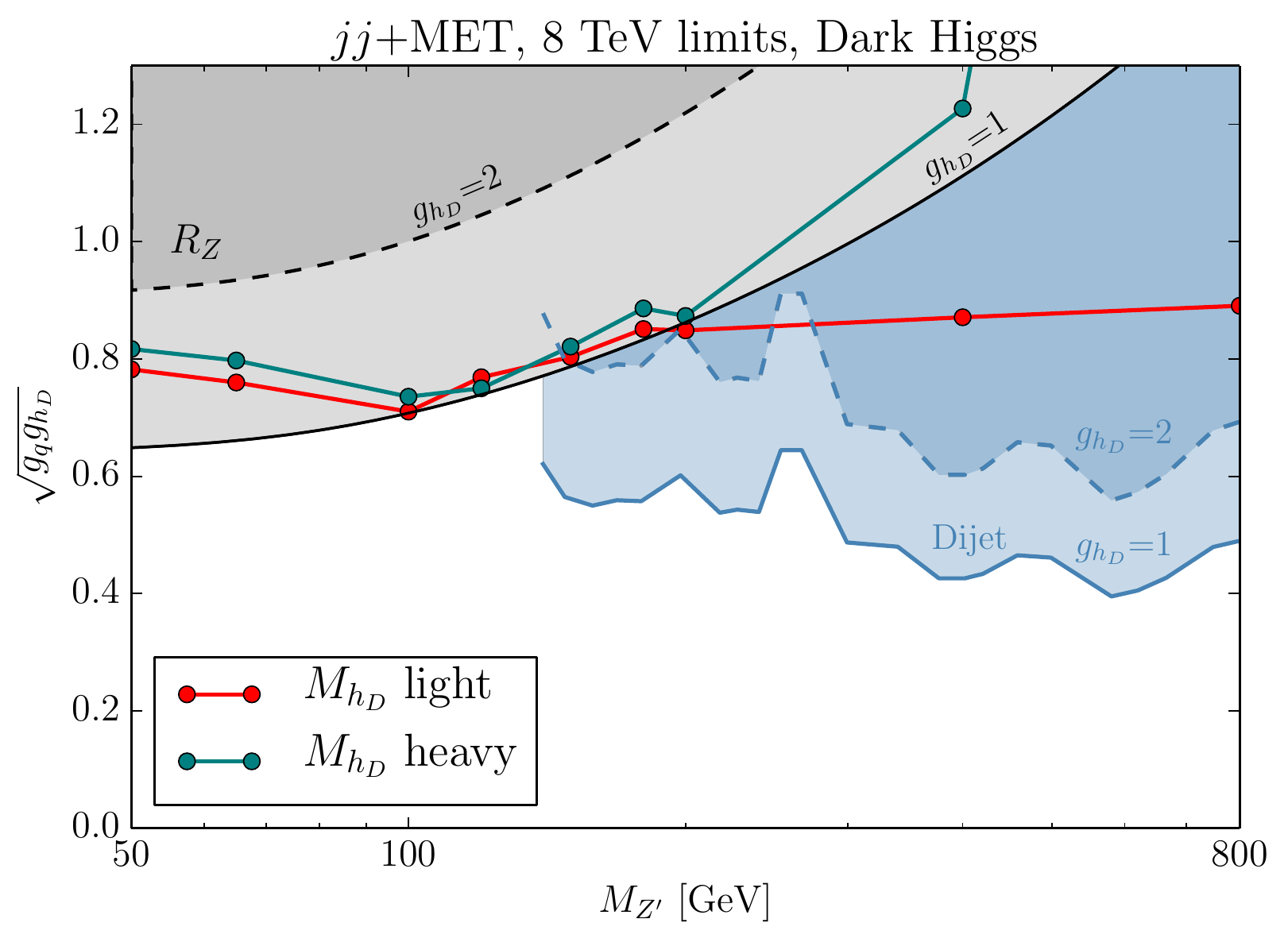}
\includegraphics[width=0.48\linewidth]{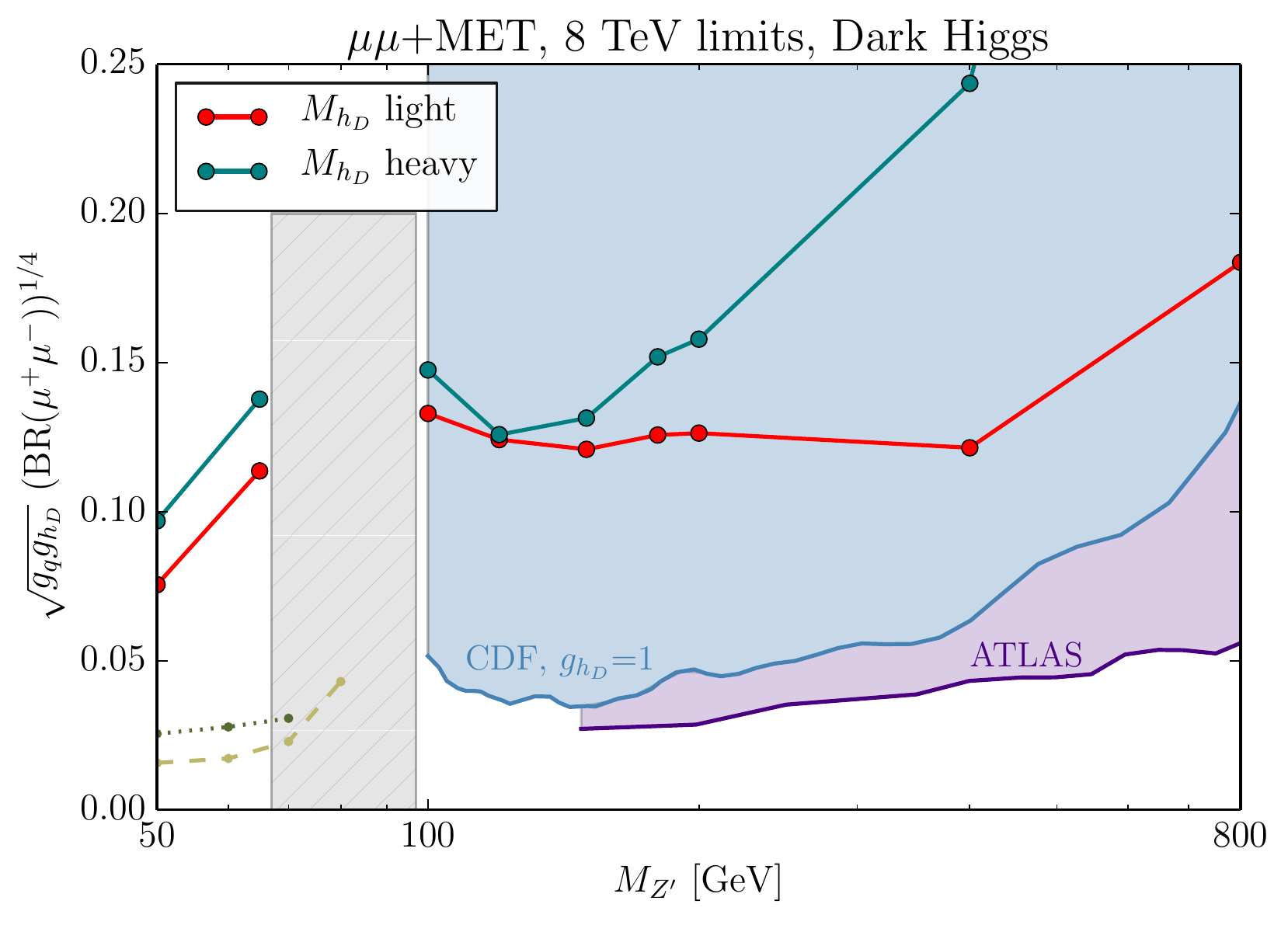}
\caption{ Expected upper limits at 95\% CL  on the product of couplings $g_q g_{h_D}$ as a function of $M_{Z'}$ for the dark Higgs model, for 8 TeV $pp$ collisions in two different mass benchmarks. 
Left, the sensitivity of the $jj + \missET$ channel is compared to the constraint on the hadronic $Z$ width (labelled $R_Z$), shown in black for $g_{h_D} = 1$ (solid) and $g_{h_D} = 2$ (dashed), as well as direct dijet resonance searches~\cite{Dobrescu:2013cmh} for a new $Z'$.  
Right, the sensitivity of the $\mu\mu + \missET$ channel is compared to various dimuon resonance searches at CDF~\cite{Aaltonen:2008ah}  and ATLAS~\cite{Aad:2014cka}, all shown for $g_{h_D} = 1$. The low-mass dimuon limits are interpreted from the results of Ref.~\cite{Hoenig:2014dsa}: both 7 TeV recast limits (dotted) and 8 TeV sensitivity projections (dashed) are shown. We do not consider masses in the grey shaded region due to the extremely large Drell-Yan background near the $Z$ mass.
\label{fig:darkHiggs_limits} }
\vspace{0.5cm}
\includegraphics[width=0.47\linewidth]{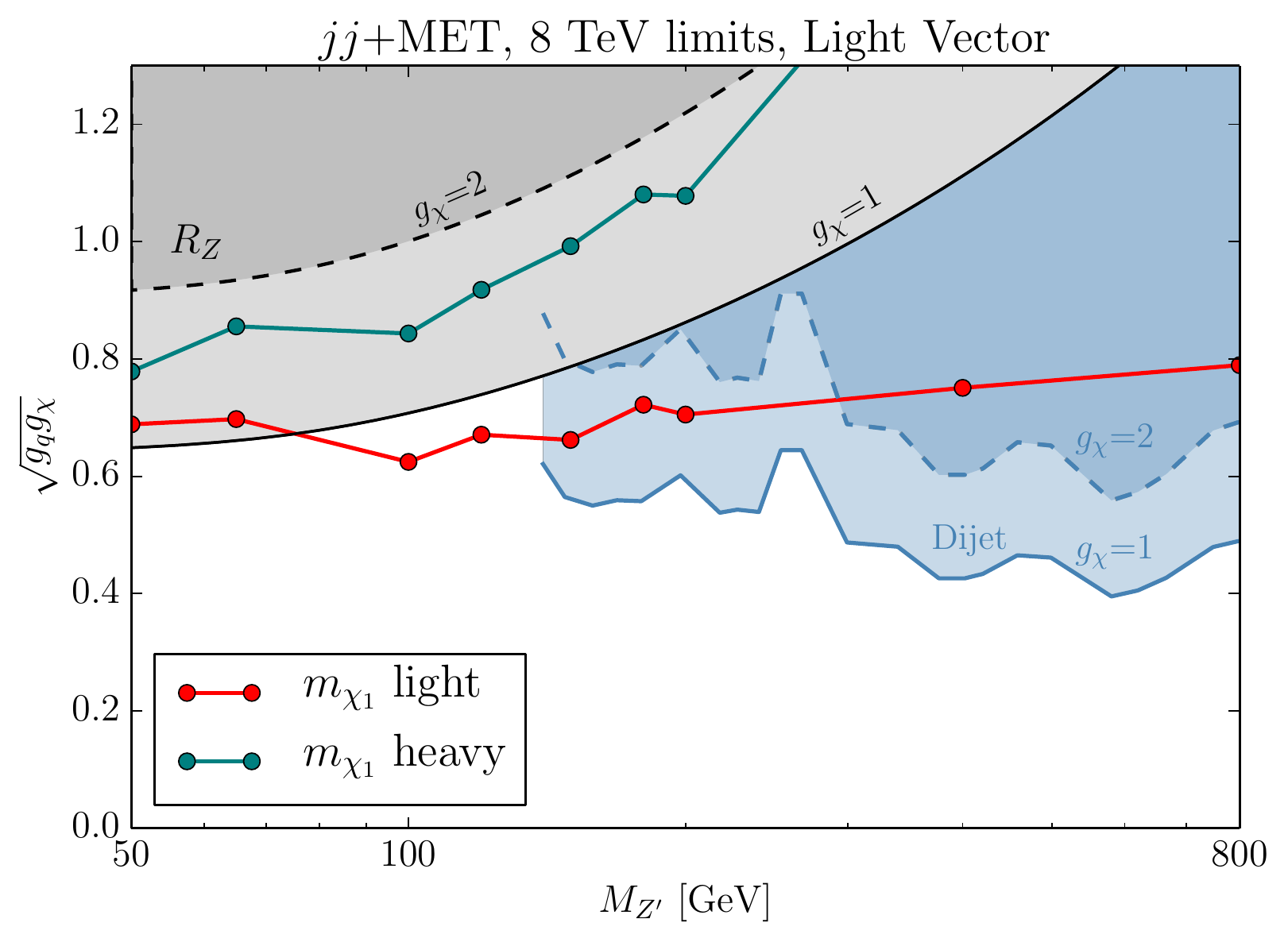}
\includegraphics[width=0.48\linewidth]{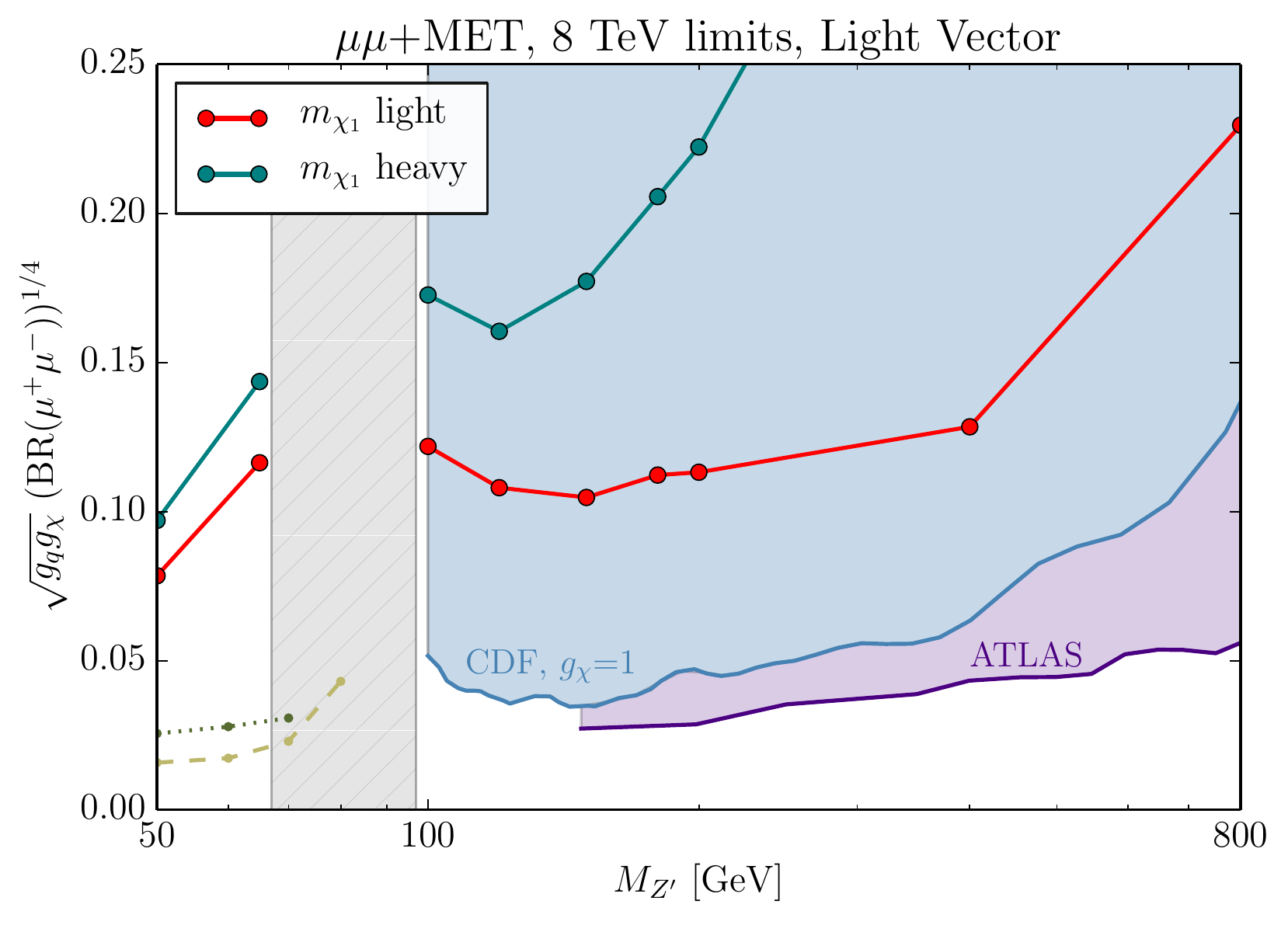}
\caption{  Expected upper limits at 95\% CL  on the product of couplings $g_q g_{h_D}$ as a function of $M_{Z'}$ in the Light Vector model, for 8 TeV $pp$ collisions in two different mass benchmarks.  The dijet and dilepton resonance limits are the same as those in Fig.~\ref{fig:darkHiggs_limits}, with $g_{\chi}=1$ for all dilepton resonance limts. \\
\ \\
\label{fig:lightVector_limits} }
\end{figure*} 
	
\begin{figure*}[t]
\begin{center}
 \includegraphics[width=0.47\linewidth]{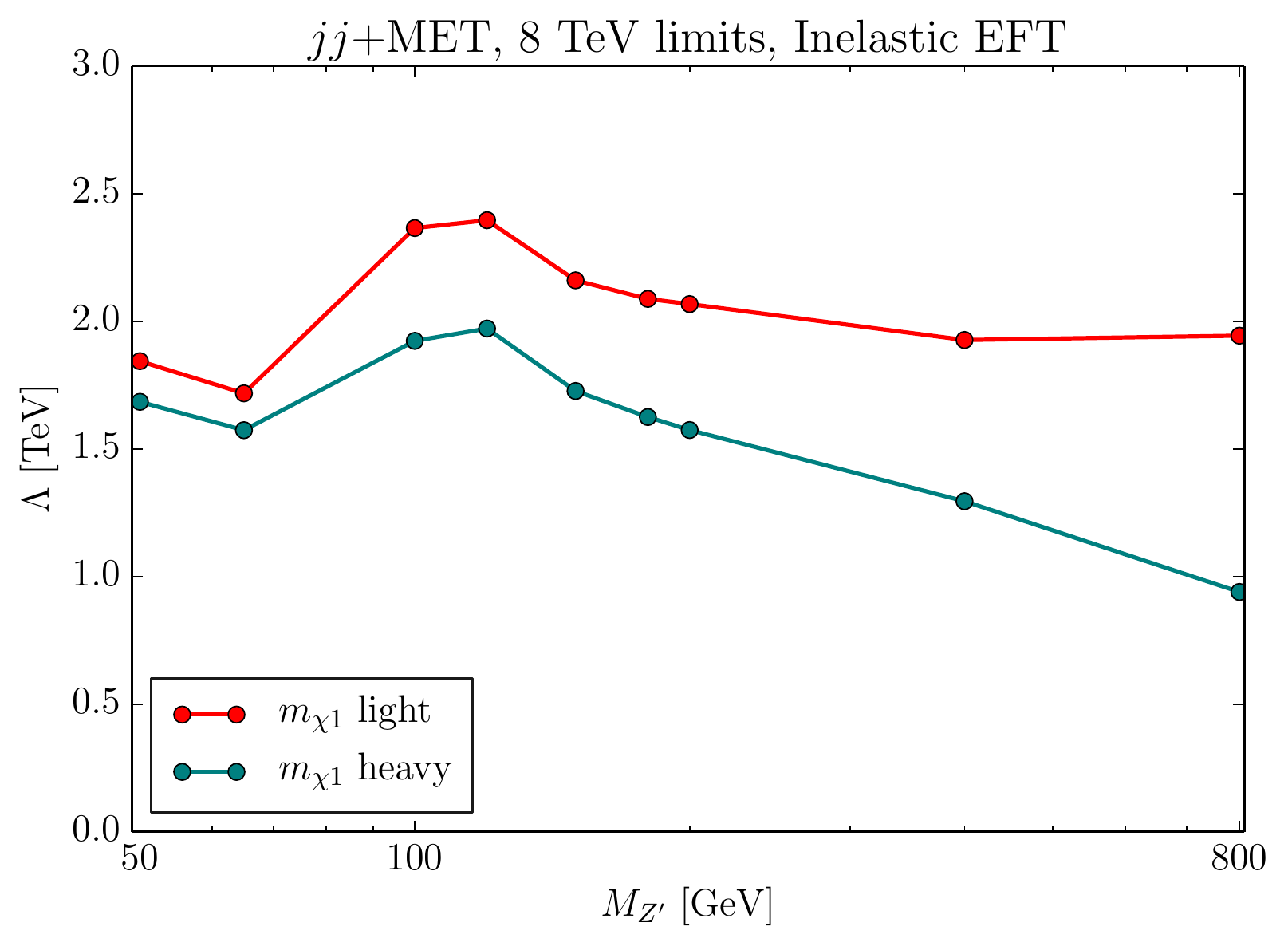}
 \includegraphics[width=0.46\linewidth]{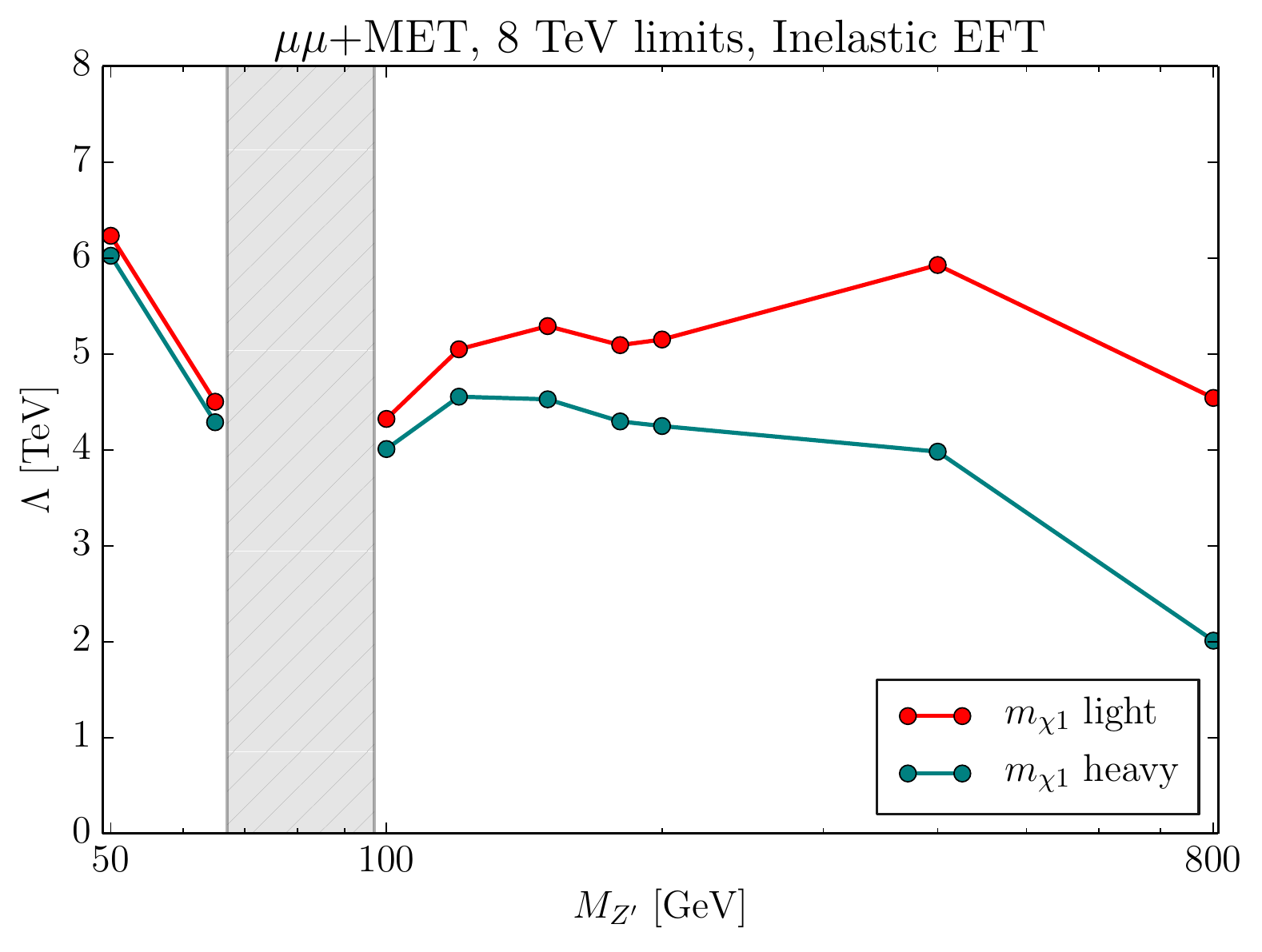}
 \caption{ Expected lower bound at 95\% CL on $\Lambda$ from in the Inelastic EFT model, for 8 TeV $pp$ collisions in two different mass benchmarks.   The branching ratio of the $Z'$ to jets and muons is taken to be 100$\%$ in each case.
 \label{fig:Inelastic_limits} }
\end{center}
\end{figure*}

The kinematic distributions in $\missET$ and invariant masses of the different models are shown in Figs.~\ref{fig:jj_kin} and \ref{fig:ll_kin}. In both the dark Higgs and light vector models, the intermediate $s-$channel $Z'$ is off-shell, and so the $\missET$ spectra are typically softer than in the inelastic EFT model and primarily determined by total mass in the final state. As a result, the high $\missET$ tail can look similar for different $Z'$ masses, if the other masses are correspondingly adjusted. Note that for the dark Higgs model, the $\missET$ spectra depends on the mass of the dark Higgs and $Z'$, and not directly on the dark matter mass, while for the light vector model the spectra depend on the total mass in the $\chi_1 \chi_2$ final state as well as on their splitting.

In the inelastic EFT model, production goes through a higher dimension operator, leading to harder $\missET$ spectra and less sensitivity to the masses in the final state. Note that the high $\missET$ tail in the $M_{Z'} = 50$ GeV case has an additional suppression, however, since such highly boosted low-mass $Z'$ are unlikely to be resolved as two individual jets. Another effect that becomes important is the size of the splitting $m_{\chi_2} - m_{\chi_1}$ compared to $M_{Z'}$: when the $\chi_1$ is very light and the splitting is very to close to $M_{Z'}$, the $p_{\textrm{T}}$ of $\chi_2$ is transferred nearly entirely to the $Z'$ and consequently the $\missET$ spectrum is harder. This corresponds to the case in Eq.~\ref{eq:spectrum1}. Conversely, less $p_{\textrm{T}}$ is transfered to the $Z'$ as the splitting is increased and as $\chi_1$ becomes heavier, as in Eq.~\ref{eq:spectrum2}.  The effect competes against the increase in missing transverse momentum with larger $m_{\chi_2}, m_{\chi_1}$. For the cases shown here, as we increase the dark matter masses, we also scale the splitting up accordingly. As a result, for a given $Z'$, the $\missET$ distribution does not change much for the two different mass spectra we consider.

\subsection{Model Constraints}

We evaluate the sensitivity of the first LHC run to each of the models presented in this paper. The results are shown in Figs.~\ref{fig:darkHiggs_limits}-\ref{fig:Inelastic_limits}, considering both dijet and dilepton resonances in the mass range $M_{Z'} = 50-800$ GeV. For each final state, we show results assuming a 100$\%$ branching ratio of the $Z'$ to dijets or to dimuons according to our signal regions.  For the dimuon final states, we do not consider the mass range $M_{Z'} \in ( 65,100)$ GeV since there is a significant Drell-Yan background at these invariant masses, as shown in Fig.~\ref{fig:ll_bg}.

As discussed in the Constraints section, there are strong constraints on electron couplings to the $Z'$, which severely limits the $Z'$ branching ratio to electrons in the dark Higgs and light vector models. For uniformity in our presentation of results we have therefore considered only the muonic final state.  The combined dimuon and di-electron result would be somewhat stronger in the case that the $Z'$ decays to both flavors equally, as in the inelastic EFT model.

Constraints for the dark Higgs model are shown in Fig.~\ref{fig:darkHiggs_limits}, for each of the two choices of dark Higgs mass given in Eqs.~\ref{eq:darkhiggslight}-\ref{eq:darkhiggsheavy}. The predicted cross section for the $Z' + \missET$ signal is proportional to $g_{h_D}^2 g_q^2$.  For $M_{Z'}<200$ GeV, the constraints for the two $M_{h_D}$ cases are similar. The lighter $M_{h_D}$ case has a larger cross section, but at the cost of a softer $\missET$ spectrum and hence reduced selection efficiency, as shown in Fig.~\ref{fig:jj_limits} and Fig.~\ref{fig:ll_limits}. Above $M_{Z'} = 200$ GeV, the limits on the heavy $M_{h_D}$ scenario become significantly weaker due to the rapidly decreasing production cross sections.

The missing transverse momentum searches are compared in each case with the corresponding direct dijet or dilepton resonance searches from various hadron colliders. Since the $Z' + \missET$ limits depend on an additional model parameter $g_{h_D}$, we show the resonance search limits for a reference value of $g_{h_D} = 1$; if this coupling were stronger, these limits would be relatively weaker. As can be seen, for $M_{Z'} > 150 (100)$ GeV in the dijet (dilepton) case, the direct resonance searches give stronger constraints on the model. 

At low $M_{Z'}$, constraints from the experimental collaborations are not available. However, we compare the dimuon results with the low mass dimuon resonance study in Ref.~\cite{Hoenig:2014dsa}, finding that their recast limits of 7 TeV data would still be stronger than that from $Z' + \missET$. Although a $\missET$ search helps reduce backgrounds, the statistics for the signal are also lower: in this model the mono-$Z'$ signal  requires an off-shell intermediate $Z'$ and the production of an additional particle (the dark Higgs) in association with the $Z'$, thus leading to a suppression of $\sim 10^3$ in the rate even for the ``light" $M_{h_D}$ case. 

We find the most relevance for this signal model in the context of leptophobic $Z'$s with mass below $\sim 150$ GeV, where there is a gap in existing dijet resonance studies. As discussed in the Constraints section, there is an indirect constraint since a light $Z'$ would modify the hadronic $Z$ width, which we show in Fig.~\ref{fig:darkHiggs_limits} for $g_{h_D}=1$ and $g_{h_D}=2$. An LHC associated $Z'$ search~\cite{An:2012ue} offers the best prospects for robust constraints competitive with the $Z' + \missET$ results in this mass range.

The limits in the light vector model are shown in Fig.~\ref{fig:lightVector_limits}, and the behavior is qualitatively similar. In addition, we make the analogous assumptions as in the dark Higgs results described above, with the resonance search results shown for $g_\chi=1$. We find that the dijet resonance plus $\missET$ performs more favorably here, having the best sensitivity to the light $m_{\chi_1}$ scenario below $M_{Z'} \approx 200-300$ GeV. However, the dimuon plus missing transverse momentum search would again be weaker than a direct dimuon search in the entire mass range.

Finally, the inelastic EFT model limits are shown in Fig.~\ref{fig:Inelastic_limits}. We constrain $\Lambda$, the scale of the operator leading to dark matter production, for each of the two channels. Since the $Z'$ can be very weakly coupled in this model, the dijet and dimuon resonance limits above do not apply and by construction, the $Z' + \missET$ search provides the best constraint. This model is especially interesting for the dimuon mode, where limits on $\Lambda$ reach roughly 5 TeV, or around 3 TeV if rescaled to ${\rm Br}(\mu^+ \mu^-) = 0.12$.

We also compare the results of our $Z'$ plus missing transverse momentum search to constraints derived using existing $\missET$-based searches for new physics beyond the standard model. For the dijet resonance plus missing transverse momentum case, the monojet search region would be sensitive to our models since up to two jets are allowed. However, by focusing on specific $m_{jj}$ windows, our analysis has far lower backgrounds. We also compare with the multijet plus missing transverse momentum SUSY search~\cite{Aad:2014wea}: we find the SUSY study is less sensitive to our models, since it requires a much larger amount of visible and missing transverse momentum in order to optimize for a signal from new heavy colored particles. In the dilepton resonance case, we compare with the chargino search~\cite{Aad:2014vma}. Here we find a fair amount of overlap in the signal regions, leading to comparable sensitivity to our models; for a more detailed discussion of the bounds obtained from applying the chargino search, see~\cite{Primulando:future}.

\section{Conclusions}

We have presented a new collider signal for dark matter: missing transverse momentum and a dijet or dilepton resonance.  This work adds to the existing mono-X and simplified models of missing transverse momentum signals, expanding the coverage of LHC searches to new dark sector physics that may be difficult to observe in other channels. In this paper, we  introduce several simplified models for a $Z'$ produced in association with the dark matter, determine the sensitivity of the current LHC dataset to these models, and compare with other collider searches for $Z'$s.

When the $Z'$ plus dark matter production relies on the $Z'$ couplings to quarks, we find that a mono-$Z'$ channel is more sensitive than dijet resonance searches only below $M_{Z'}$ of a few hundred GeV. In this mass range, there are currently no published results searching for a resonance from a hadronically decaying $Z'$, and the requirement of $\missET$ can significantly reduce the QCD dijet background.  On the other hand, in these same models, when the $Z'$ can also decay to leptons then a direct dilepton resonance search is expected to be a more powerful constraint on the model in the entire $Z'$ mass range.

The $Z'$ can also be produced in the decay of dark sector states, which are coupled to quarks through an effective contact interaction. Then the $Z'$ may be weakly coupled to SM states, easily satisfying other direct collider constraints. Such a model would be challenging to observe in other missing transverse momentum searches, but give rise to a mono-$Z'$ signal. As the first run of the LHC has shown, there is need for a broad range of dark matter signals to explore the many possibilities for the dark sector and to take full advantage of the data.

\subsection{Acknowledgements}
We thank Yang Bai, Gordan Krnjaic, Reinard Primulando, Jessie Shelton, Tim Tait, Liantao Wang, Felix Yu, and Ning Zhou for useful discussions, and we are especially grateful to Prashant Saraswat for helpful comments on this draft. This work was supported in part by the Kavli Institute for Cosmological Physics at the University of Chicago through grant NSF PHY-1125897 and an endowment from the Kavli Foundation and its founder Fred Kavli. TL thanks the Center for Future High Energy Physics (CFHEP) in Beijing for hospitality and partial support.

\bibliography{monozp}

\clearpage
\appendix

\end{document}